\documentclass[12pt]{article}
\usepackage{amsmath,amsfonts,amsthm,amssymb}
\usepackage{graphicx,psfrag,epsf}
\usepackage{enumerate}
\usepackage{natbib}
\usepackage{url} %
\usepackage{xr-hyper} 
\usepackage[pagewise]{lineno}

\usepackage{setspace}
\usepackage[english]{babel}
\usepackage{Tabbing}
\usepackage{fancyhdr}
\usepackage{lastpage}
\usepackage{extramarks}
\usepackage{chngpage}
\usepackage{soul,color}
\usepackage{graphicx,float,wrapfig}
\usepackage{layout}

\usepackage{latexsym}
\usepackage{amsmath}
\usepackage{multirow}
\usepackage{amssymb}
\usepackage{amsbsy}
\usepackage{amsthm}
\usepackage{amsfonts}
\usepackage{mathrsfs}
\usepackage{mathtools}
\usepackage{bm}
\usepackage{relsize}
\usepackage{caption2}
\usepackage{graphicx}
\usepackage{subfigure}
\usepackage{array}
\usepackage{natbib}
\usepackage{placeins}
\usepackage[colorlinks,linkcolor=blue,anchorcolor=blue,citecolor=blue]{hyperref}

\newcommand{\blind}{0}

\newtheorem{thm}{Theorem}
\newtheorem{rem}{Remark}
\newtheorem{lem}[thm]{Lemma}

\makeatletter
\def\singlespace{\def\baselinestretch{1}\@normalsize}

\makeatletter
\def\singlespace{\def\baselinestretch{1}\@normalsize}

\numberwithin{equation}{section}

\renewcommand{\hat}{\widehat}

\newcommand{\balpha}{\bm{\alpha}}

\newcommand{\bphi}{\bm{\phi}}

\newcommand{\bal}{\mbox{\boldmath$\alpha$}}

\newcommand{\btheta}{\bm{\theta}}
\newcommand{\bTheta}{\bm{\Theta}}

\newcommand{\eps}{\varepsilon}

\def\eps{\varepsilon}

\def\newpage{\vfill\eject}

\def\today{\ifcase\month\or
	January\or February\or March\or April\or May\or June\or
	July\or August\or September\or October\or November\or December\fi
	\space\number\day, \number\year}

\def\no{\noindent}

\newdimen\biblioindent    \biblioindent=30pt

 at 8truept

\def\balpha{\bal}

\def\eps{\varepsilon}

\newcommand{\beq}{\begin{equation}}
	\newcommand{\eeq}{\end{equation}}
\newcommand{\beqn}{\begin{eqnarray}}
	\newcommand{\eeqn}{\end{eqnarray}}
\newcommand{\beqnn}{\begin{eqnarray*}}
	\newcommand{\eeqnn}{\end{eqnarray*}}

\DeclareMathOperator{\E}{\mathbb{E}}

\DeclareMathOperator{\Var}{\mbox{Var}}

\providecommand{\abs}[1]{\left\lvert#1\right\rvert}

\begin{document}

	\def\spacingset#1{\renewcommand{\baselinestretch}%
		{#1}\small\normalsize} \spacingset{1}

	\if0\blind
	{
		\title{\bf  Robust Estimation of Double Autoregressive Models via Normal Mixture QMLE}
		\author{ \normalsize Zhao Chen\footnote{Zhao Chen is Professor, School of Data Science, Fudan University, Shanghai, China 200433. Email: zchen\_fdu@fudan.edu.cn. } \\ \vspace{-2in} \small Fudan University   \and \normalsize Chen Shi\footnote{Chen Shi is Ph.D. Candidate, School of Data Science, Fudan University, Shanghai, China 200433. Email: cshi22@m.fudan.edu.cn.} \\ \vspace{-2in} \small Fudan University     \and  \normalsize Christina Dan Wang \footnote{ Corresponding author: Christina Dan Wang is Assistant Professor, Business Division, New York University Shanghai, Shanghai, China 200122. Email: christina.wang@nyu.edu. }\\ \small New York University Shanghai}
		
		\maketitle
	} \fi
	
	\bigskip
	
	\begin{abstract}
	This paper investigates the estimation of the double autoregressive (DAR) model in the presence of skewed and heavy-tailed innovations. We propose a novel Normal Mixture Quasi-Maximum Likelihood Estimation (NM-QMLE) method to address the limitations of conventional quasi-maximum likelihood estimation (QMLE) under non-Gaussian conditions. By incorporating a normal mixture distribution into the quasi-likelihood framework, NM-QMLE effectively captures both heavy-tailed behavior and skewness. A critical contribution of this paper is addressing the often-overlooked challenge of selecting the appropriate number of mixture components, $K$, a key parameter that significantly impacts model performance. We systematically evaluate the effectiveness of different model selection criteria. Under regularity conditions, we establish the consistency and asymptotic normality of the NM-QMLE estimator for DAR(p) models. Numerical simulations demonstrate that NM-QMLE outperforms commonly adopted QMLE methods in terms of estimation accuracy, particularly when the innovation distribution deviates from normality. Our results also show that while criteria like BIC and ICL improve parameter estimation of $K$, fixing a small order of components provides comparable accuracy. To further validate its practical applicability, we apply NM-QMLE to empirical data from the S\&P 500 index and assess its performance through Value at Risk (VaR) estimation. The empirical findings highlight the effectiveness of NM-QMLE in modeling real-world financial data and improving risk assessment. By providing a robust and flexible estimation approach, NM-QMLE enhances the analysis of time series models with complex innovation structures, making it a valuable tool in econometrics and financial modeling.
\end{abstract}

\noindent%
{\it Keywords:}  Heavy-tailed, Skewed, Normal mixture, Quasi-maximum likelihood estimation
\vfill

\newpage
\spacingset{1.8} %
\section{Introduction}
\label{sec:intro}

The double autoregressive~(DAR) model is an important framework in time series analysis, particularly for capturing conditional heteroscedasticity in economic and financial data. It can be considered as a special case of ARMA-ARCH model proposed by \cite{W1986} and a weak ARMA model studied by~\cite{FZ1998, FZ2000}.  \cite{L2004, L2007} investigated the stationarity properties of DAR($1$) and DAR($p$) models and established the asymptotic
theory of Gaussian maximum likelihood estimation~(GMLE). However, in financial and econometric applications, the Gaussian assumption on the underlying distribution of innovations is often not satisfied. When the true innovation distribution is unknown, Gaussian quasi maximum likelihood estimation~(GQMLE)
is commonly employed due to its simplicity, such as in \cite{FZ2010} and others. Theoretically, GQMLE remains robust and consistent under the assumption of a finite fourth moment for the innovations, as discussed in \cite{W1986, BHK2003} and \cite{BH2004}. However, when the underlying distribution
deviates significantly from normality, the estimation efficiency of GQMLE deteriorates and the variance of the
estimate increases.

Although GQMLE is a feasible approach in various applications, empirical studies have shown unsatisfactory performance
when applied to data exhibiting non-normal features, see for example \cite{MS2000} and \cite{HY2003}. To address the limitation of GQMLE, the existing literature has considered more flexible quasi-likelihood functions that can capture non-normal behavior, such as skewness or heavy tails in innovations.
For example, non-Gaussian QMLE has been adopted to estimate the GARCH type of models, , including estimators based on exponential density \cite{BH2004} and double exponential density \cite{ZL2011}. Furthermore, \cite{GL2020} proposed a multi-steps Student's t-QMLE.  However, while these approaches primarily address heavy-tailed behavior, they often fall short when the innovation distribution also exhibits asymmetry. To incorporate the skewness of the innovation distribution, the normal mixture model has been proposed as an alternative.

The normal mixture model, as discussed by \cite{MP2000}, is  well-suited for capturing both skewness and heavy-tailed behavior. In a GARCH framework, \cite{R2001} and \cite{BRT2003} considered two component normal mixture models as innovation distribution, while \cite{HMP2004} proposed generalized normal mixture GARCH models.  Subsequent extensions by \cite{LA2004} and \cite{ZLY2006} refined these models further by employing symmetric normal mixture GARCH models. Extending this idea, \cite{LL2009} introduced the normal mixture QMLE for GARCH models, \cite{HL2011} applied it to ARMA-GARCH models, and \cite{WP2014} utilized it for non-stationary TGARCH models. These works demonstrated the utility of normal mixture QMLE in capturing complex data structures. %

However, despite these advances, a key limitation persists across this body of work: the number of mixture components $K$ is typically fixed—often arbitrarily—without theoretical or empirical justification. The literature offers little guidance on how to select $K$ appropriately, leaving a critical gap in model specification and practical implementation. This omission can affect both the accuracy of parameter estimates and the interpretability of results. Moreover, these mixture-based models often depart from the standard GARCH framework, introducing additional complexity that hinders asymptotic inference. A similar concern applies when imposing normal mixture distributions on DAR models, as doing so can alter their fundamental structure in undesirable ways.

To address these challenges, this paper seeks to make two key contributions. First, we propose a novel Normal Mixture Quasi-Maximum Likelihood Estimation (NM-QMLE) method for DAR($p$) models. This method preserves the original DAR model structure while leveraging the flexibility of normal mixture distributions. Unlike traditional GQMLE, NM-QMLE is designed to capture both heavy-tailed behavior and skewness in the data, ensuring more reliable parameter estimates under complex innovation structures. Second, we systematically address the practical challenge of selecting the order of components in the normal mixture distribution. By analyzing criteria such as the Bayesian Information Criterion (BIC) and Integrated Classification Likelihood (ICL), we demonstrate that ICL achieves superior accuracy in identifying the true component structure, while fixing a small order provides a computationally efficient alternative without sacrificing much accuracy.

On the theoretical front, we derive the asymptotic properties of the NM-QMLE estimator under mild assumptions. Specifically, we establish the consistency of the estimator and demonstrate its asymptotic normality, providing a solid theoretical foundation for its application in scenarios where the innovation distribution is unknown or significantly non-normal. Complementing our theoretical results, we perform extensive Monte Carlo simulations to assess the finite-sample performance of the NM-QMLE. These simulation studies focus on DAR(1) and DAR(2) models under conditions where the innovations exhibit heavy-tailed and skewed characteristics. The simulation results clearly indicate that the NM-QMLE method achieves lower estimation error compared to existing QMLE methods, thereby validating its superior performance in capturing the underlying dynamics of the data. 

In addition, we conduct a detailed analysis of the selection of the number of mixture components ($K$) in the NM-QMLE method. We compare the performance of different criteria including AIC, BIC, ICL, and the slope heuristic, in various simulation settings. Based on the simulation results, we suggest that ICL provides the most accurate estimate results. For practical use, we find that fixing a small value for $K$ yielding reliable parameter estimates with lower computational cost, especially when the true distribution is unknown or potentially misspecified.

Additionally, we apply the NM-QMLE method to an empirical analysis using daily returns of the S\&P 500 index, covering the period from January 3, 2007, to December 29, 2023. Since the true return distribution is not directly observable, we evaluate the practical performance of our estimator through its ability to generate accurate Value at Risk (VaR) estimates---a widely used risk measure in financial markets that quantifies the potential loss over a specified time horizon at a given confidence level. The empirical findings demonstrate that the VaR estimates based on NM-QMLE are statistically valid and robust, underscoring the method's practical utility in financial risk management and its effectiveness in capturing tail behavior.

Overall, by integrating normal mixture distributions into the quasi-likelihood framework, our proposed NM-QMLE method not only preserves the structural integrity of the DAR model but also offers significant improvements in estimation accuracy and inference. This work provides a comprehensive and flexible tool for researchers and practitioners dealing with non-normal innovations in time series analysis, particularly in the context of financial econometrics.

\section{Methodology}
\label{sec:meth}
\subsection{Model Setup}
Consider the DAR($p$) model for a time series $\{y_{t} \}$:
\begin{equation}\label{Mod:DAR}
	\begin{split}
		& {y}_{t} = \phi_1 {y}_{t-1} + \cdots + \phi_p y_{t-p} + \eps_{t},  \\
		&  \eps_{t} = \eta_{t} \sqrt{\omega + \alpha_1 {y}_{t-1}^{2} + \cdots + \alpha_p {y}_{t-p}^{2}} \, ,
	\end{split}
\end{equation}
where $\bphi = (\phi_1,\ldots, \phi_p)^T, \balpha = (\alpha_1, \ldots, \alpha_p)^T$,
and $\btheta_{1} = (\bphi^T, \omega, \balpha^T)^T$ are the unknown parameters. Model~\eqref{Mod:DAR} requires $\omega, \alpha_1, \ldots, \alpha_p >0$. Assume that the parameter space $\bTheta_{1} \subset (-\infty, \infty)^p \times (0, \infty)^{p+1}$ is a compact set and the true parameters $\btheta_{10} = (\bphi^{0^T}, \omega^{0}, \balpha^{0^T})^{T}$
is an interior point in $\bTheta_{1}$.
The innovation $\{\eta_{t}, -\infty < t < \infty \}$ are i.i.d. random variables with the density function $g(\cdot)$ and satisfy $\E \eta_{t} = 0 \text{ and var}(\eta_{t}) = 1$. The initial $y_{0}$ is usually assumed to be independent of $\{ \eta_{t}, t \geq 1\}$.

In this paper, we assume that $\{y_t\}$ in model~\eqref{Mod:DAR} is strictly stationary and ergodic.
The existence of a stationary solution to DAR model is still an open question. \cite{L2007}
established the necessary and sufficient conditions for the stationary solution with normal innovation.
Suppose $\{\xi_{j t}\}, j=1, \ldots, p$, are i.i.d. standard normal random variables and
are independent of the innovation $\{\eta_t\}$. Let
\beq
A_t = \begin{pmatrix}
	\phi_1 + \sqrt{\alpha_1} \xi_{1t} & \cdots &  \phi_{p-1} + \sqrt{\alpha_{p-1}} \xi_{p-1,t}
	&  \phi_p + \sqrt{\alpha_p} \xi_{pt}  \\
	& & & \\
	& I_{p-1} & &  \bm{0}_{(p-1) \times 1} \\
	& & &
\end{pmatrix},
\eeq
where $I_{p-1}$ is a $(p-1)\times (p-1) $ identity matrix. \cite{L2007} showed that
$\{y_t\}$ with the finite second moment is a strictly stationary solution to DAR($p$) model
if and only if that all the eigenvalues of $\E A_t \otimes A_t$ lie between -1 and 1.
For the simplest case $p=1$, the aforementioned condition reduces to $\phi_1^2 + \alpha_1 < 1$,
which is the same as the condition in \cite{GD1994}
and  \cite{BK1998}. Therefore, we assume that
$\bTheta_{1} =
\{ (\bphi^T, \omega, \balpha^T)^{T}: \rho\left( \E A_t \otimes A_t \right) <1,
\ \abs{\phi_j} \leq \widetilde{\phi}_j,\ \underline{\omega} \leq \omega \leq \overline{\omega},\
\underline{\alpha_j} \leq \alpha_j \leq \overline{\alpha_j},\ j = 1, \ldots, p \}$,
where  $\rho(A)$ is the spectral radius of the matrix $A$ (i.e. $\rho(A)=\max_i|\nu_i|$, where $\nu_i$ 's are the eigenvalues of $A$), and $\widetilde{\phi}_j$, $\underline{\omega}$, $\overline{\omega}$, $\underline{\alpha}_j$,
$\overline{\alpha}_j$ are all positive constants.

\subsection{Normal Mixture Quasi-Maximum Likelihood Estimation }

If the innovation is assumed to be normally distributed, the parameters can be estimated by GMLE. Even if the specific distribution of innovation is unknown, normal or double exponential quasi-maximum likelihood estimation~(QMLE) can still be considered. However, such methods become less effective when the true distribution function deviates significantly from the quasi-likelihood function. In this paper, the normal mixture quasi-maximum likelihood estimate is proposed to improve the cases with general unknown distribution functions of innovations.

First, we introduce the family of normal mixture densities. Suppose the $\sigma$-field generated by the normal mixture density functions is
\begin{eqnarray*}
	\mathscr{F} = \{ g_{\btheta_{2}}: \btheta_{2} \in \bTheta_{2} \}.
\end{eqnarray*}
Here $g_{\btheta_{2}}$ is a weighted average of $K$ different normal densities,
\beq\label{g_theta2}
g_{\btheta_{2}}(y) = \sum_{k=1}^{K} p_{k} f(y; \mu_{k}, \sigma_{k}),
\eeq
where $K$ is a fixed positive integer, $f(y; \mu_{k}, \sigma_k)$'s are the normal density functions with mean $\mu_k$
variance $\sigma_k^2$,
\begin{eqnarray}
	f(y; \mu_{k}, \sigma_{k}) =
	\frac{1}{\sqrt{2\pi} \sigma_{k}}
	\exp \displaystyle\bigg\{ -\frac{(y - \mu_{k})^{2}}{2 \sigma_{k}^{2}} \displaystyle\bigg\}, \quad \sigma_k >0,
\end{eqnarray}
and $p_k$ are the weights satisfying
\beqn
& & \sum_{k=1}^{K} p_{k} = 1,  \label{Theta2_1} \\
& & \sum_{k=1}^{K} p_{k}\mu_{k} = 0,  \label{Theta2_2} \\
& & \sum_{k=1}^{K} p_{k}(\mu_{k}^{2} + \sigma_{k}^{2}) = 1.  \label{Theta2_3}
\eeqn
The family $\mathscr{F}$ of $K$-component normal mixture
density functions is usually not identifiable. To obtain a well-defined parameter estimation, it is usually
assumed that the $\sigma$-field $\mathscr{F}$ is weakly identifiable, that is,

\medskip

\no\textbf{Assumption (Identifiability, ID)}\label{ID} {\it
	$\mathscr{F}$ is identifiable, if the following condition holds:
	\begin{eqnarray*}
		\sum_{k=1}^{K} p_{k}^{1} f(y; \mu_{k}^{1}, \sigma_{k}^{1}) =
		\sum_{k=1}^{K} p_{k}^{2} f(y; \mu_{k}^{2}, \sigma_{k}^{2}) \
		\xLeftrightarrow{\mbox{ ~a.e. }} \
		\sum_{k=1}^{K} p_{k}^{1} \delta_{(\mu_{k}^{1}, \sigma_{k}^{1})}(a, b) =
		\sum_{k=1}^{K} p_{k}^{2} \delta_{(\mu_{k}^{2}, \sigma_{k}^{2})}(a, b),
	\end{eqnarray*}
	where $\delta_{(a, b)}(\cdot)$ satisfies:
	$\delta_{(a,b)}(a,b) = 1$ and $\delta_{(a, b)}(x,y) = 0$, for any $(x,y) \neq (a,b)$. }

\medskip

\noindent For convenience, denote
\begin{eqnarray*}
	\btheta_{2} = (p_{1}, p_{2}, \cdots, p_{K-1};
	\mu_{1}, \mu_{2}, \cdots, \mu_{K-1};
	\sigma_{1}, \sigma_{2}, \cdots, \sigma_{K-1})^{T}.
\end{eqnarray*}
The parameter space $\bTheta_{2} \subset [0,1]^{K-1} \times \mathbb{R}^{K-1} \times \mathbb{R}_{+}^{K-1}$ satisfies the constraints \eqref{Theta2_1}-\eqref{Theta2_3}. To introduce the normal mixture quasi-maximum likelihood estimation, we further need the following definition.

\medskip

\noindent\textbf{Definition~1.} {\it
	For a given density function $g$, the Kullback-Leibler~(KL) divergence functional is defined as follows:
	\beq\label{KLdiver}
	\mathscr{T}(g) = \{ \btheta_{2} \in \bTheta_{2}: d(g, g_{\btheta_{2}}) = \min_{t \in \bTheta_{2}} d(g, g_{t}) \},
	\eeq
	where
	\begin{eqnarray*}
		d(g, g_{t}) = \int g(y) (\log g(y) - \log g_{t}(y)) d \nu(y).
	\end{eqnarray*}
	is the commonly used KL divergence.}

\medskip

Recall that the innovation has the density function $g(\cdot)$ and the normal mixture density is $g_{\btheta_2}(\cdot)$. According to the definition of KL divergence functional,
we set the true parameter of $\btheta_{2}$  as $\btheta_{20} \in \mathscr{T}(g)$,
\begin{eqnarray*}
	\btheta_{20} =
	(p_{1}^{0}, p_{2}^{0}, \cdots, p_{K-1}^{0};
	\mu_{1}^{0}, \mu_{2}^{0}, \cdots, \mu_{K-1}^{0};
	\sigma_{1}^{0}, \sigma_{2}^{0}, \cdots, \sigma_{K-1}^{0})^{T}.
\end{eqnarray*}
Suppose the observed sample $\{y_{1}, y_{2}, \cdots, y_{n}\}$ is generated by model~\eqref{Mod:DAR}. We replace
the unknown density function with normal mixture density as a working likelihood function. Given the initial value $y_{0}$, the conditional normal mixture quasi-likelihood function is defined as
\begin{align}
	L_{n}(\btheta) = \prod_{t=1}^{n}
	\displaystyle\bigg\{
	\sum_{k=1}^{K}
	&p_{k} \frac{1}{\sqrt{2 \pi \sigma_{k}^{2} h_t(\omega, \balpha)}}  \nonumber \\
	&
	\exp \displaystyle\bigg[
	-\frac{(y_{t} - m_{t}(\bphi) - \mu_{k} \sqrt{h_t(\omega, \balpha)})^{2}}
	{2 \sigma_{k}^{2} h_t(\omega, \balpha)}
	\displaystyle\bigg]
	\displaystyle\bigg\},
\end{align}
where $m_{t}(\bphi) = \phi_1 y_{t-1} + \cdots + \phi_p y_{t-p}$ and
$h_t(\omega, \balpha) = \omega + \alpha_1 y_{t-1}^2 + \cdots + \alpha_p y_{t-p}^2$.
Redefine the unknown parameters of interest as
\begin{eqnarray*}
	\btheta = (\btheta_{1}^{T}, \btheta_{2}^{T})^{T}
	= (\bphi^T, \omega, \balpha^T;
	p_{1}, \cdots, p_{K-1},
	\mu_{1}, \cdots, \mu_{K-1},
	\sigma_{1}, \cdots, \sigma_{K-1})^{T}.
\end{eqnarray*}
The corresponding true value is denoted by $\btheta_{0} = (\btheta_{10}^{T}, \btheta_{20}^{T})^{T}$
and the parameter space is $\bTheta = \bTheta_{1} \times \bTheta_{2}$.
Finally, the normal mixture quasi maximal likelihood estimator~(NMQMLE) of $\btheta_{0}$ is
obtained by maximizing the log-quasi-likelihood, that is
\begin{eqnarray}
	\widehat{\btheta}_{n} = \mathop{\arg\max}\limits_{\btheta \in \bTheta} L_{n}(\btheta) =
	\mathop{\arg\max}\limits_{\btheta \in \bTheta} \log L_{n}(\btheta) \triangleq
	\mathop{\arg\min}\limits_{\btheta \in \bTheta} - l_{n}(\btheta),
\end{eqnarray}
where
\beq
l_{n}(\btheta) = \frac{1}{n} \sum_{t=1}^{n} W_{t}(\btheta),
\eeq
and
\begin{align}
	W_{t}(\btheta)
	&= \log \displaystyle\bigg\{
	\sum_{k=1}^{K}
	p_{k} \frac{1}{\sqrt{2 \pi \sigma_{k}^{2} h_t(\omega, \balpha)}}  \nonumber  \\
	& \quad \quad \exp \displaystyle\bigg(
	-\frac{(y_{t} - m_{t}(\bphi) - \mu_{k} \sqrt{h_t(\omega, \balpha)})^{2}}
	{2 \sigma_{k}^{2} h_t(\omega, \balpha)}
	\displaystyle\bigg)
	\displaystyle\bigg\}  \nonumber  \\
	&= \log \displaystyle\bigg\{
	\frac{1}{\sqrt{h_t(\omega, \balpha)}}
	g_{\btheta_{2}}
	\displaystyle\bigg(
	\frac{y_{t} - m_{t}(\bphi)}{\sqrt{h_t(\omega, \balpha)}}
	\displaystyle\bigg)
	\displaystyle\bigg\},   \quad g_{\btheta_{2}} \in \mathscr{F}. \label{WtDef}
\end{align}

\section{Determining the number of components in NMQMLE}

A critical issue when applying NM-QMLE to estimate DAR($p$) model is the selection of the number of mixture components $K$, which corresponds to the order of the underlying mixture distribution. One classical approach to determining $K$ is through formal hypothesis testing using the likelihood ratio test (LRT). However, as has been extensively discussed in~\citet{MP2000,MR2014,MLR2019}, the standard regularity conditions do not hold for the null distribution of the likelihood-ratio test statistic to have its usual chi-squared distribution with degrees of freedom equal to the difference between the number of parameters under the null and alternative hypotheses.  Consequently, alternative strategies such as penalized likelihood criteria are often preferred in practice for determining the number of components. There are a number of well-known methods to select the best mixture model but the Bayesian information criterion
\cite{S1978} remains by far and away the most popular.
\begin{align}
	\mathrm{BIC}(K) = -2 \log L_n(\widehat{\btheta}_{n}) + D_K \log n,
\end{align}
where $D_K = 2p+1 + 3(K-1)$ is the number of free parameters under $K$ components, and $n$ is the number of observations.

From a theoretical standpoint, \cite{K2000} demonstrated that BIC is a consistent estimator of the number of components in finite mixture models under certain regularity conditions. However, in practice, it has often been observed that BIC tends to overestimate the number of components \citep{BCG2002}. To address this limitation, \cite{BCG2002} proposed the Integrated Classification Likelihood (ICL) criterion, which augments the BIC by incorporating an entropy penalty term that accounts for the uncertainty in component classification.
\begin{align}
	ICL(K) = -2 \log L_n(\widehat{\btheta}_{n}) + D_K \log n  - 2 \sum_{i=1}^n\sum_{k=1}^K \hat{\tau}_{ik} \log \hat{\tau}_{ik},
\end{align}
where $\hat{\tau}_{ik}$ denotes the posterior probability that observation $i$ originates from component $k$, and is computed as
\begin{align*}
	\hat{\tau}_{ik}=\frac{\hat{p}_k \cdot \mathcal{N}\left(\hat{\eta}_t  \mid \hat{\mu}_k, \hat{\sigma}_k^2\right)}{\sum_{j=1}^K \hat{p}_j \cdot \mathcal{N}\left(\hat{\eta}_t  \mid \hat{\mu}_j, \hat{\sigma}_j^2\right)}, \quad \hat{\eta}_t = \frac{y_{t} - m_{t}(\hat{\boldsymbol{\phi}})}{\sqrt{h_t(\hat{\omega}, \hat{\boldsymbol{\alpha}})}}.
\end{align*}

Another refinement of the BIC that we considered is the slope heuristics method proposed by \cite{BM2007}. They argued that the  $\log n$ penalty coefficient in the BIC may be too conservative in practice and suggested a data-driven approach to estimate the appropriate penalty constant based on the shape of the log-likelihood curve. Practical implementation details and algorithmic procedures are elaborated in \cite{BMM2012}. In Section \ref{sec:sim}, we conduct simulation experiments to evaluate the performance of different model selection criteria in recovering the true number of mixture components and in guiding the estimation of DAR($p$) model parameters when the innovation distribution is unknown. We assess both the selection accuracy of the true number of components and the estimation accuracy of the autoregressive parameters under quasi-maximum likelihood estimation (QMLE) using mixture normal approximations.

Our findings suggest that the ICL criterion achieves the highest accuracy in identifying the true number of components, followed by BIC, while slope heuristics performs less reliably. Furthermore, in the more realistic setting where the true innovation density is unknown, we show that selecting $K$ via ICL leads to more accurate DAR($p$) model parameter estimates. Nevertheless, we also find that precise identification of the true number of components is not strictly necessary for effective QMLE estimation. In particular, we observe that fixing a small value of $K$, such as $K=2$, yields estimators with satisfactory accuracy while offering improved robustness and significantly reduced computational burden. Although model selection criteria like BIC and ICL can, in some cases, lead to marginal improvements in estimation precision, these gains often come at the cost of substantially greater computational complexity. As illustrated by our simulation results, adopting a modest, fixed value of $K$ is often a pragmatic and effective strategy for QMLE estimation in DAR models.

\section{Asymptotic properties of ~NMQMLE}
In this section, we study the asymptotic properties of NMQMLE. The following assumptions are imposed to
facilitate the proof and are adopted from several existing literature~\citep{L2007, LL2009, WP2014, GL2020}.
They may not be the
weakest possible conditions. %
Suppose the following assumptions hold:

\bigskip

\no\textbf{Assumption A1.} The parameter space $\bTheta $ is a compact set. The true $\btheta_0$
is an interior point of the parameter space $\bTheta$.

\no\textbf{Assumption A2.}
$\E y_t^2 < \infty$ and $\{ y_{t}, t = 0,1, \cdots\}$ is strictly stationary and ergodic.

\no\textbf{Assumption A3.} The innovation $\{\eta_t\}$ is an i.i.d. standard white noise process and
has density function $g(\cdot)$.

\no\textbf{Assumption A4.}
$\btheta_{20} \in \bTheta_{2}$ is essentially unique and satisfies that
\begin{enumerate}
	\item[(a)] when $\btheta_{2} \neq \btheta_{20}$,
	there exists
	\begin{eqnarray*}
		\E \log \{ u g_{\btheta_{2}}(u \eta_{t}) \} < \E \log g_{\btheta_{20}}(\eta_t)
		\quad \mbox{for any } u > 0 .
	\end{eqnarray*}
	\item[(b)] for any $a \in \mathbb{R}$, $\E \log g_{\btheta_{20}}(a \eta_{t} + b)$ attains its maxima
	at $b = 0$.
\end{enumerate}

\no\textbf{Assumption ~A5.}
$\E \eta_{t}^{4} < + \infty$.

\no\textbf{Assumption ~A6.}
The following matrices~$H$ and~$J$ are non-singular:
\begin{eqnarray}
	H = \E \displaystyle\bigg\{ - \frac{\partial^{2}W_{t}(\btheta_{0}) }{ \partial \btheta \partial \btheta^T}  \displaystyle\bigg\},\quad
	J = \Var \displaystyle\bigg\{ \frac{\partial W_{t}(\btheta_{0})}{\partial \btheta} \displaystyle\bigg\}.  \label{HandJ}
\end{eqnarray}

\vspace{0.1in}\

\no With these assumptions, we establish the asymptotic properties of NMQMLE.

\begin{thm}
	\label{theorem41}
	Under assumptions ~A1-A4, it follows that
	\begin{eqnarray}
		\widehat{\btheta}_{n} \xrightarrow{P} \btheta_{0}, \quad \text{ as } n \rightarrow \infty.
	\end{eqnarray}
\end{thm}

\begin{thm}
	\label{theorem42}
	Under assumptions ~A1-A6 and ~(ID), it follows that
	\begin{eqnarray}
		\sqrt{n}(
		\widehat{\btheta}_{n} - \btheta_{0} )
		\xrightarrow{d}
		N \displaystyle\bigg(0, H^{-1} J H^{-1} \displaystyle\bigg), \text{ as } n \rightarrow \infty, \label{normaliseAll}
	\end{eqnarray}
	where ~$H$ and ~$J$ refer to equation (\ref{HandJ}).
\end{thm}

\begin{rem}
	In general, $\mathscr{F}$, the family of linear combinations of $K$ normal density functions,  is not identifiable. Therefore, we have to impose  the condition (ID) and redefine the the parameter space to achieve identifiability.
	See \cite{R1984} for more details.
\end{rem}

\begin{rem}
	Assumption A4 is adapted from \cite{newey1997asymptotic}, which provides a location and scale identification condition necessary for the consistency of non-Gaussian QMLE. This type of identification restriction plays a crucial role in ensuring that the objective function has a unique maximizer at the pseudo-true value, especially under distributional misspecification. Subsequent studies, such as \cite{fiorentini2007efficiency,fiorentini2019consistent}, propose consistent estimators to replace those parameters that are inconsistently estimated by misspecified log-likelihoods in multivariate, conditionally heteroskedastic, dynamic regression settings. 
\end{rem}

\begin{rem} As stated in \cite{LL2009}, if $\,d(g, g_{\btheta_{2}})$ is a continuous function with respect to parameter $\btheta_{2}$, then the maxima $\btheta_{20}$ exists. Therefore, by transformation of parameters, the parameter space for normal mixture densities can be embedded into a compact set.
\end{rem}

\section{Numerical studies} \label{sec:sim}
In this section, we conduct a series of Monte Carlo simulations to evaluate the performance of NM-QMLE for DAR models. We examine three key aspects: (i) the estimation accuracy of NM-QMLE under various innovation distributions with a fixed number of components; (ii) the effectiveness of different criteria in selecting the true number of mixture components when the innovations follow a finite normal mixture; and (iii) the impact of different criteria selection of $K$  on parameter estimation accuracy when the true innovation distribution is unknown and possibly misspecified.

\subsection{Estimation Performance under Fixed $K$}

We first examine its finite-sample performance of QMLE under DAR(1) and DAR(2) models with heavy-tailed and skewed innovation distributions, which pose significant challenges and currently lack effective estimation tools. To demonstrate the advantages of our approach, we compare NM-QMLE with the conventional QMLE and the recently proposed three-step non-Gaussian quasi-maximum likelihood estimator (TS-NGQMLE) by \cite{GL2020}. While the asymptotic properties of standard QMLE are well documented in \cite{L2004}, the theoretical properties of TS-NGQMLE are discussed in \cite{GL2020}. In this experiment, we fix the number of mixture components at $K=2$, which serves as a simple yet flexible approximation for a variety of innovation distributions.

Given that \cite{GL2020} has already demonstrated the superiority of TS-NGQMLE over other existing methods, we focus our comparison to the classical QMLE and TS-NGQMLE. Specifically, we evaluate the estimation accuracy of the parameter $\theta = (\phi, \omega, \alpha)^{T}$ and compare the  performance among our approach NM-QMLE, QMLE and TS-MGQMLE.

The simulation study considers both DAR(1) and DAR(2) models. For DAR(1), the parameters $\theta_{0} = (\phi_{0}, \omega_{0}, \alpha_{0})^{T}$ are set as $(0.3, 1.0, 0.5)^{T}$ or $(0.0, 1.0, 0.9)^{T}$. For DAR(2), the parameters $\theta_{1} = (\phi_{1}, \phi_2, \omega_{0}, \alpha_{1}, \alpha_2)^{T}$ are set as $(0.3, 0.1, 1.0, 0.5, 0.2)^{T}$ or $(0.0, 0.1, 1.0, 0.9, 0.2)^{T}$.

\begin{figure}[!hbp]
	\centering
	\includegraphics[width=0.7\textwidth]{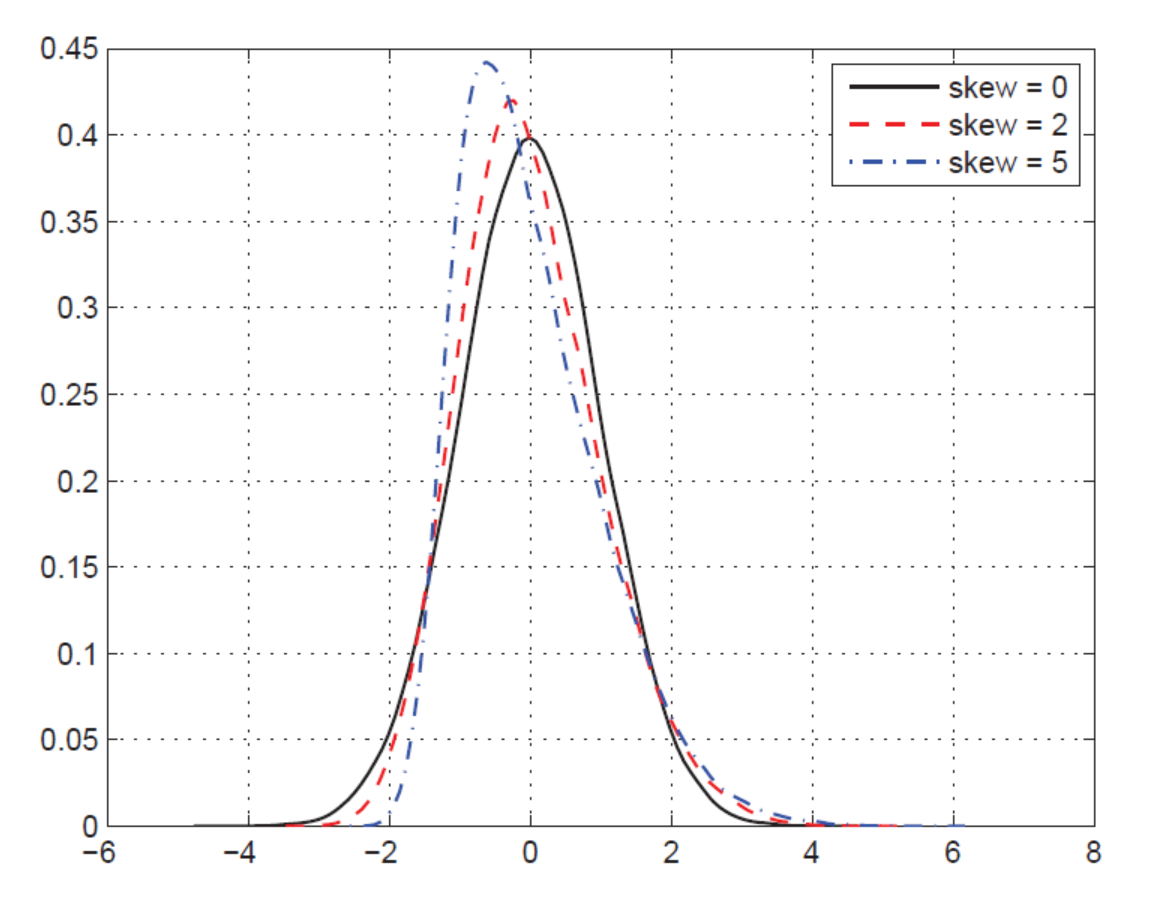}
	\caption{Different skewed normal distributions under different parameters.}           \label{fig41}
\end{figure}

We consider 9 different scenarios for the innovation distribution $g$ of $\eta$. They are
\begin{itemize}
	\item[(a)]student t distribution with degree of freedom (df) 2.5;
	\item[(b)]student t distribution with $df=5$;
	\item[(c)] student t distribution with $df=10$;
	\item[(d)] skew normal distribution with parameter $\theta= 2$;
	\item[(e)] skew normal distribution with parameter $\theta= 5$;
	\item[(f)] skew normal distribution with parameter $\theta= 10$.
	\item[(g)] skewed t distribution with parameters $q=2.5,\ \lambda= -0.9$;
	\item[(h)] skewed t distribution with parameters $q=4,\ \lambda= -0.5$;
	\item[(i)] skewed t distribution with parameters $q=2.5,\ \lambda= 0.3$.
\end{itemize}
For each scenario, we replicate 1000 sample paths with sample size set as $n = 1000$.

\begin{rem}
	\label{remark41}
	For the simulation of the skew normal distribution, we follow the approach outlined in \cite{GLD2014}. For convenience, Appendix 1 provides a detailed description of the general simulation method. To offer an intuitive understanding, Figure \ref{fig41} illustrates the density functions of the skew normal distribution under various parameter settings.
\end{rem}

The simulation results for the DAR(1) model are presented in Table \ref{tab:1} and Table \ref{tab:2}, with $ \theta_{0} = (\phi_{0}, \omega_{0}, \alpha_{0})^{T}$ taking values ~$(0.3, 1.0, 0.5)^{T}$ and $(0.0, 1.0, 0.9)^{T}$, respectively. Since the simulations employ t-distributed and skewed normal innovations, the true MLE can be computed as a benchmark reference. 

As expected, the root mean squared error (RMSE) of the true MLE is the smallest across all scenarios. Almost for all models under different parameterizations, the newly proposed NM-QMLE outperforms Gaussian QMLE and TS-NGQMLE except one scenario where innovations follow a $t_5$ distribution.  In this specific case, TS-NGQMLE achieves slightly better performance than NM-QMLE because the third step of TS-NGQMLE employs a $t_5$ distribution, which coincides with the true innovation distribution.  This explains why TS-NGQMLE performs almost as well as the true MLE. Even for this special case, NM-QMLE only performs slightly worse in terms of errors. These results demonstrate the effectiveness of NM-QMLE in improving estimation accuracy when dealing with skewed and heavy-tailed distributions.

For skewed distributions, the advantage of NM-QMLE over other methods becomes more pronounced. When the skewness is moderate, such as in the case of a skewed normal distribution with parameter $\theta=2$, NM-QMLE achieves the best performance, although the other two methods also perform reasonably well. And it is worth notice that in this scenario, QMLE may outperform the TS-NGQMLE. However, skewness becomes more severe, the performance of QMLE and TS-NGQMLE deteriorates rapidly, whereas NM-QMLE continues to behave closely to the true MLE. In these cases, TS-NGQMLE performs slightly better than QMLE, but neither matches the accuracy of NM-QMLE.  

For skewed and heavy-tailed distributions, NM-QMLE consistently outperforms other estimation methods across all scenarios. Its advantage becomes particularly evident as the distribution exhibits heavier tails, especially in cases where the tail index parameter is  $p=2.5$. Notably, NM-QMLE maintains superior performance regardless of whether the skewness is positive or negative, demonstrating its effectiveness in handling complex distributional features.

Table \ref{tab:3} and Table \ref{tab:4} present the simulation results for the DAR(2) model. The findings are consistent with those observed in the DAR(1) case. Specifically, for skewed and heavy-tailed distributions, our proposed method, NM-QMLE, significantly outperforms TS-NGQMLE.

\begin{frame}
	\scriptsize
	\begin{table}[htbp]
		\centering
		\caption{RMSE of estimation with different methods for $(0.3, 1.0, 0.5)^{T}$ for DAR($1$) and $n=1000$.}
		\begin{tabular}{c|lrrrr}\hline\hline
			&       & \multicolumn{1}{c}{MLE} & \multicolumn{1}{c}{QMLE} & \multicolumn{1}{c}{TS-NGQMLE} & \multicolumn{1}{c}{NM-QMLE} \\\hline
			\multirow{3}[0]{*}{$t_{2.5}$} & $\phi$ & 0.0296 & 0.1081 & 0.0326 & 0.0317 \\
			& $\omega$ & 0.0860 & 1.5565 & 0.8653 & 0.5098 \\
			& $\alpha$ & 0.1328 & 0.6424 & 0.2447 & 0.2478 \\
			\hline
			\multirow{3}[0]{*}{$t_{5}$} & $\phi$ & 0.0364 & 0.0407 & 0.0366 & 0.0369 \\
			& $\omega$ & 0.0872 & 0.1268 & 0.1055 & 0.1087 \\
			& $\alpha$ & 0.0824 & 0.1116 & 0.0875 & 0.0922 \\
			\hline
			\multirow{3}[0]{*}{$t_{10}$} & $\phi$ & 0.0385 & 0.0393 & 0.0387 & 0.0385 \\
			& $\omega$ & 0.0785 & 0.0846 & 0.0817 & 0.0813 \\
			& $\alpha$ & 0.0729 & 0.0774 & 0.0735 & 0.0735 \\\hline
			
			\multirow{3}[0]{*}{skew normal(2)} & $\phi$ & 0.0383 & 0.0398 & 0.0414 & 0.0388 \\
			& $\omega$ & 0.0667 & 0.0764 & 0.0991 & 0.0691 \\
			& $\alpha$ & 0.0580 & 0.0655 & 0.0722 & 0.0592 \\       \hline
			
			\multirow{3}[0]{*}{skew normal(5)} & $\phi$ &  0.0303  & 0.0400 & 0.0436 & 0.0326 \\
			& $\omega$ & 0.0597 & 0.0859 & 0.1032 & 0.0693 \\
			& $\alpha$ & 0.0512 & 0.0755 & 0.0741 & 0.0575 \\       \hline
			
			\multirow{3}[0]{*}{skew normal(10)} & $\phi$ & 0.0224 & 0.0384 & 0.0436 & 0.0282 \\
			& $\omega$ &  0.0454 & 0.0871 & 0.1010 & 0.0641 \\
			& $\alpha$ & 0.0373 & 0.0747 & 0.0710 & 0.0511 \\ \hline
			\multirow{3}[0]{*}{skewed t(2.5, -0.9)} & $\phi$ & 0.0111 & 0.0566 & 0.0780 & 0.0297 \\
			& $\omega$ &  0.0226 & 1.1736 & 0.8525 & 0.5512 \\
			& $\alpha$ & 0.0496  & 1.5753 & 0.4625 & 0.4296 \\ \hline
			\multirow{3}[0]{*}{skewed t(4, -0.5)} & $\phi$ & 0.0300  & 0.0452 & 0.0446 & 0.0341 \\
			& $\omega$ &  0.0652 & 0.2039 & 0.1736 & 0.1560 \\
			& $\alpha$ & 0.0701  &  0.3508 & 0.1121 & 0.1118 \\ \hline
			\multirow{3}[0]{*}{skewed t(2.5, 0.3)} & $\phi$ & 0.0275  & 0.0505 &  0.0369  & 0.0301 \\
			& $\omega$ &   0.0746 & 2.8497 & 1.7575 & 0.4238 \\
			& $\alpha$ & 0.1185   &  1.0023  & 0.7710  & 0.2562 \\\hline \hline
		\end{tabular}%
		\label{tab:1}%
	\end{table}%
\end{frame}

\begin{table}[htbp]
	\centering
	\caption{ RMSE of estimation with different methods for $(0.0, 1.0, 0.9)^{T}$ for DAR($1$) and $n=1000$.}
	\begin{tabular}{c|lrrrr}\hline\hline
		&       & \multicolumn{1}{c}{MLE} & \multicolumn{1}{c}{QMLE} & \multicolumn{1}{c}{TS-NGQMLE} & \multicolumn{1}{c}{NM-QMLE} \\ \hline
		\multirow{3}[0]{*}{$t_{2.5}$} & $\phi$ & 0.0351 & 0.0589 & 0.0361 & 0.0345 \\
		& $\omega$ & 0.0882     & 0.3698        & 0.3489       &   0.3217 \\
		& $\alpha$ & 0.1887    &     0.8184      &    0.3483   &    0.3459 \\
		\hline
		\multirow{3}[0]{*}{$t_5$} & $\phi$ & 0.0406 & 0.0462 & 0.0406 & 0.0414 \\
		& $\omega$ & 0.0933 & 0.1334 & 0.1389 & 0.1153 \\
		& $\alpha$ & 0.1083 & 0.1500 & 0.1440 & 0.1202 \\
		\hline
		\multirow{3}[0]{*}{$t_{10}$} & $\phi$ & 0.0448 & 0.0455 & 0.0454 & 0.0446 \\
		& $\omega$ & 0.0878 & 0.0957 & 0.1146 & 0.0916 \\
		& $\alpha$ & 0.0962 & 0.1039 & 0.1235 & 0.0991 \\
		\hline
		\multirow{3}[0]{*}{skew normal(2)} & $\phi$ & 0.0423 & 0.0434 & 0.0446 & 0.0426 \\
		& $\omega$ & 0.0759 & 0.0851 & 0.0842 & 0.0789 \\
		& $\alpha$ & 0.0786 & 0.0860 & 0.0869 & 0.0805 \\
		\hline
		\multirow{3}[0]{*}{skew normal(5)} & $\phi$ & 0.0331 & 0.0428 & 0.0447 & 0.0354 \\
		& $\omega$ & 0.0658 & 0.0979 & 0.0910 & 0.0772 \\
		& $\alpha$ & 0.0629 & 0.0883 & 0.0823 & 0.0705 \\
		\hline
		\multirow{3}[0]{*}{skew normal(10)} & $\phi$ & 0.0255 & 0.0442 & 0.0474 & 0.0323 \\
		& $\omega$ & 0.0516 & 0.0973 & 0.0881 & 0.0696 \\
		& $\alpha$ & 0.0491 & 0.0943 & 0.0844 & 0.0685 \\ \hline

		\multirow{3}[0]{*}{skewed t(2.5, -0.9)} & $\phi$ &  0.0120 &  0.0608 & 0.0905 & 0.0303 \\
		& $\omega$ & 0.0241 & 0.7398 & 0.5342 & 0.3595 \\
		& $\alpha$ & 0.0628 & 1.4756 & 0.4622 & 0.3312 \\ \hline
		\multirow{3}[0]{*}{skewed t(4, -0.5)} & $\phi$ & 0.0326  & 0.0487  &  0.0528  & 0.0371 \\
		& $\omega$ &  0.0741 & 0.2305 & 0.1838 & 0.1528 \\
		& $\alpha$ & 0.0945  &  0.4542 & 0.1815 & 0.1209 \\ \hline
		\multirow{3}[0]{*}{skewed t(2.5, 0.3)} & $\phi$ & 0.0312  & 0.0547 &  0.0420  &  0.0342 \\
		& $\omega$ &   0.0766 & 2.6485 & 1.4991 & 0.3085 \\
		& $\alpha$ & 0.1690   & 1.9661 & 1.1513 & 0.2996 \\ \hline  \hline
	\end{tabular}%
	\label{tab:2}%
\end{table}%

\begin{table}[htbp]
	\centering
	\renewcommand{\arraystretch}{0.97}
	\caption{ RMSE of estimation with different methods for $(0.3, 0.1, 1.0, 0.5, 0.2)^{T}$ for DAR($2$) and $n=1000$.}
	\begin{tabular}{c|lrrrr}\hline\hline
		&       & \multicolumn{1}{c}{MLE} & \multicolumn{1}{c}{QMLE} & \multicolumn{1}{c}{TS-NGQMLE} & \multicolumn{1}{c}{NM-QMLE} \\ \hline
		\multirow{5}[0]{*}{$t_{2.5}$} & $\phi_1$ & 0.0310 & 0.0852 & 0.0335 & 0.0328 \\
		& $\phi_2$ & 0.0281 & 0.0579 & 0.0310 & 0.0299 \\
		& $\omega$ & 0.0944     & 0.5278       & 0.3487       &   0.3099\\
		& $\alpha_1$ & 0.1375   &     0.5477     &    0.1969  &   0.1959 \\
		& $\alpha_2$ & 0.0918    &     0.4935    &   0.1051   &    0.0999 \\
		\hline
		\multirow{5}[0]{*}{$t_5$} & $\phi_1$ & 0.0379 & 0.0427 & 0.0381 & 0.0385 \\
		& $\phi_2$ & 0.0334 & 0.0380 & 0.0336& 0.0337 \\
		& $\omega$ & 0.1086 & 0.1453& 0.1176 & 0.1216 \\
		& $\alpha_1$ &0.0853   &    0.1191    &    0.0890  &   0.0930 \\
		& $\alpha_2$ &0.0571    &     0.0763    &   0.0578  &  0.0613 \\
		\hline
		\multirow{5}[0]{*}{$t_{10}$} & $\phi_1$ & 0.0403 &0.0418 & 0.0403 & 0.0408 \\
		& $\phi_2$ & 0.0362 &0.0373& 0.0362 & 0.0366 \\
		& $\omega$ & 0.1050 &0.1105 &0.1055 &0.1058 \\
		& $\alpha_1$ & 0.0727   &     0.0761   &    0.0731  & 0.0737\\
		& $\alpha_2$ &0.0482   &     0.0526    &  0.0484  &    0.0498 \\
		\hline
		\multirow{5}[0]{*}{skew normal(2)} & $\phi_1$ &0.0366& 0.0389 & 0.0395 & 0.0372 \\
		& $\phi_2$ & 0.0339 & 0.0358 & 0.0366 &0.0344 \\
		& $\omega$ & 0.0909 & 0.0999& 0.0986 & 0.0933 \\
		& $\alpha_1$ &0.0642   & 0.0694     &  0.0697  &  0.0649 \\
		& $\alpha_2$ & 0.0454   &    0.0481    &  0.0478   &   0.0459 \\
		\hline
		\multirow{5}[0]{*}{skew normal(5)} & $\phi_1$ & 0.0290 & 0.0391& 0.0407 & 0.0313 \\
		& $\phi_2$ & 0.0269 & 0.0351 &0.0363 & 0.0296 \\
		& $\omega$ & 0.0758 & 0.1111 & 0.1032 & 0.0868 \\
		& $\alpha_1$ &0.0507  & 0.0737    &   0.0682  &   0.0560 \\
		& $\alpha_2$ & 0.0381   &  0.0531   &  0.0471   &0.0417 \\
		\hline
		\multirow{5}[0]{*}{skew normal(10)} & $\phi_1$ &0.0223 & 0.0393 & 0.0416 & 0.0282 \\
		& $\phi_2$ & 0.0209 &0.0353& 0.0369 & 0.0269 \\
		& $\omega$ &0.0586 & 0.1153 & 0.1037 & 0.0817 \\
		& $\alpha_1$ & 0.0388   &   0.0761     &  0.0676  &  0.0510 \\
		& $\alpha_2$ & 0.0291    & 0.0546   &   0.0467  &    0.0373 \\
		\hline
		\multirow{5}[0]{*}{skewed t(2.5, -0.9)} & $\phi_1$ &  0.0285 &  0.0958 & 0.0364 & 0.0307 \\
		& $\phi_2$ & 0.0255 &0.1013 & 0.0310 & 0.0280 \\
		& $\omega$ & 0.0889 &1.8488  & 0.3802 & 0.3246 \\
		& $\alpha_1$ &0.1292  &  0.5778   & 0.2221  & 0.2129 \\
		& $\alpha_2$ &0.0848  &  0.5304   & 0.1151   & 0.1036 \\ 
		\hline
		\multirow{5}[0]{*}{skewed t(4, -0.5)} & $\phi_1$ &0.0378  & 0.0450 &  0.0387 & 0.0385 \\
		& $\phi_2$ & 0.0333 & 0.0391& 0.0340 &0.0338 \\
		& $\omega$ &  0.1061 &0.2164 & 0.1482 &0.1431 \\
		& $\alpha_1$ &0.0878   & 0.1633    &0.1074 &  0.1082 \\
		& $\alpha_2$ & 0.0613  &  0.1209   &0.0648   &0.0673 \\ 
		\hline
		\multirow{5}[0]{*}{skewed t(2.5, 0.3)} & $\phi_1$ &0.0305  &0.0776&  0.0336 &  0.0317 \\
		& $\phi_2$ & 0.0272 & 0.0557 & 0.0290 &0.0286 \\
		& $\omega$ &  0.0964 &0.5160 & 0.3663 &0.3048 \\
		& $\alpha_1$ & 0.1362   &0.5434     &0.2156  &0.1922 \\
		& $\alpha_2$ & 0.0933   &0.5237 & 0.1119  &0.1065 \\  \hline  \hline
	\end{tabular}%
	\label{tab:3}%
\end{table}%

\begin{table}[htbp]
	\centering
	\renewcommand{\arraystretch}{0.97}
	\caption{ RMSE of estimation with different methods for $(0, 0.1, 1.0, 0.9, 0.2)^{T}$ for DAR($2$) and $n=1000$.}
	\begin{tabular}{c|lrrrr}\hline\hline
		&       & \multicolumn{1}{c}{MLE} & \multicolumn{1}{c}{QMLE} & \multicolumn{1}{c}{TS-NGQMLE} & \multicolumn{1}{c}{NM-QMLE} \\ \hline
		\multirow{5}[0]{*}{$t_{2.5}$} & $\phi_1$ & 0.0341 & 0.0479 & 0.0357 &0.0356\\
		& $\phi_2$ & 0.0281 & 0.0545 & 0.0300 & 0.0297 \\
		& $\omega$ & 0.1011     & 0.5940       &0.3982       &   0.3123\\
		& $\alpha_1$ & 0.1925   &  0.8895    & 0.3814 &  0.3231\\
		& $\alpha_2$ & 0.0936   &  0.4673    &0.1172   & 0.1007 \\
		\hline
		\multirow{5}[0]{*}{$t_5$} & $\phi_1$ &0.0425 &0.0474 &0.0427 &0.0431 \\
		& $\phi_2$ & 0.0338 & 0.0381 & 0.0340 &0.0341 \\
		& $\omega$ & 0.1209 &0.1622 &0.1306 & 0.1339 \\
		& $\alpha_1$ &0.1188   &0.1672    &0.1284  &  0.1333 \\
		& $\alpha_2$ &0.0580   & 0.0780   &0.0589  & 0.0621 \\
		\hline
		\multirow{5}[0]{*}{$t_{10}$} & $\phi_1$ & 0.0447 &0.0463& 0.0447 &0.0452 \\
		& $\phi_2$ & 0.0360 & 0.0370 &0.0360 & 0.0365 \\
		& $\omega$ & 0.1172 & 0.1244 & 0.1174 & 0.1186 \\
		& $\alpha_1$ & 0.0985  &   0.1025   & 0.0999  &0.1007 \\
		& $\alpha_2$ & 0.0462    & 0.0500   &  0.0465  &  0.0479 \\
		\hline
		\multirow{5}[0]{*}{skew normal(2)} & $\phi_1$ & 0.0406 & 0.0431 & 0.0437 & 0.0412 \\
		& $\phi_2$ & 0.0338 &0.0357 & 0.0366 & 0.0343 \\
		& $\omega$ & 0.1032 & 0.1134 & 0.1115 & 0.1056 \\
		& $\alpha_1$ & 0.0842   & 0.0913     &0.0911  & 0.0856 \\
		& $\alpha_2$ & 0.0437  &  0.0471   &0.0461  & 0.0439 \\
		\hline
		\multirow{5}[0]{*}{skew normal(5)} & $\phi_1$ & 0.0320 &0.0429 &0.0451 &0.0344  \\
		& $\phi_2$ & 0.0270 & 0.0352 & 0.0373 & 0.0295 \\
		& $\omega$ & 0.0878 & 0.1264 &0.1180 & 0.0982 \\
		& $\alpha_1$ & 0.0657  & 0.0948    & 0.0885 &0.0738 \\
		& $\alpha_2$ & 0.0357   & 0.0500    &0.0444  & 0.0389 \\
		\hline
		\multirow{5}[0]{*}{skew normal(10)} & $\phi_1$ & 0.0245 &0.0432 & 0.0465 & 0.0307 \\
		& $\phi_2$ &0.0208 & 0.0353& 0.0382 & 0.0268 \\
		& $\omega$ & 0.0686 &0.1324 &0.1200 & 0.0917 \\
		& $\alpha_1$ & 0.0496   &0.0973     & 0.0879  & 0.0673 \\
		& $\alpha_2$ & 0.0279    & 0.0511    & 0.0444  & 0.0347 \\
		\hline
		\multirow{5}[0]{*}{skewed t(2.5, -0.9)} & $\phi_1$ &  0.0309 &  0.0463 & 0.0382 & 0.0333 \\
		& $\phi_2$ & 0.0261 & 0.1053 & 0.0325 & 0.0291 \\
		& $\omega$ & 0.0935 &1.3917 & 0.3966 & 0.3251 \\
		& $\alpha_1$ & 0.1832  &  0.9780     & 0.3987 & 0.3474  \\
		& $\alpha_2$ & 0.0870  &  0.4969  & 0.1221  & 0.1056 \\ 
		\hline
		\multirow{5}[0]{*}{skewed t(4, -0.5)} & $\phi_1$ & 0.0418 & 0.0492  &  0.0428  &0.0426 \\
		& $\phi_2$ & 0.0334 & 0.0390 & 0.0342 &0.0341 \\
		& $\omega$ & 0.1150 &0.2285 &0.1502 &0.1490 \\
		& $\alpha_1$ & 0.1214  & 0.2436     & 0.1586  & 0.1602 \\
		& $\alpha_2$ & 0.0626    & 0.1320    & 0.0660  & 0.0691 \\ 
		\hline
		\multirow{5}[0]{*}{skewed t(2.5, 0.3)} & $\phi_1$ &0.0333 &0.0487 &0.0353 &0.0348 \\
		& $\phi_2$ & 0.0276 &0.0572 & 0.0304 &0.0290 \\
		& $\omega$ &  0.1016&0.5605 & 0.3720 & 0.3053  \\
		& $\alpha_1$ & 0.1980  & 0.8603    &0.3534  & 0.3199 \\
		& $\alpha_2$ & 0.0919    & 0.4720   & 0.1099  & 0.1077 \\  \hline  \hline
	\end{tabular}%
	\label{tab:4}%
\end{table}%

\FloatBarrier

\subsection{Selection of the Number of Components via Information Criteria}
In this section, we evaluate the performance of several model selection criteria---AIC, BIC, ICL, and two slope heuristic-based methods (DDSE and Djump)---in determining the number of mixture components $K$ for the innovation distribution in the DAR model. The slope-based methods follow the approach of \cite{BMM2012} and are implemented using their R package \texttt{capushe}.

To conduct the analysis, we simulate data under a DAR(2) model where the innovation terms $\eta_t$ follow a finite mixture of normal distributions with $K_{\text{true}} = 2,3,4,5$.  Each mixture is constructed with equally weighted components, component means equally spaced between $-K_{\text{true}}$ and $K_{\text{true}}$ with component variances 0.5. After generating random samples from this Gaussian mixture, we normalize the innovations to have zero mean and unit variance. The DAR(2) model parameters are set as $\theta_{1} = (\phi_{1}, \phi_2, \omega_{0}, \alpha_{1}, \alpha_2)^{T}=(0.3, 0.1, 1.0, 0.5, 0.2)^{T}$. For each setting of $K_{\text{true}}$, we conduct 100 independent simulations with $n=1000$. In each simulation, we estimate models with $K=2,\dots,15$ mixture components and apply the above selection criteria to choose the optimal number of components $\hat{K}$. 

Table \ref{tab:select_K_freq} reports the selection frequencies of each method for different true number of components $K_{\text{true}} = 2,3,4,5$. We find that ICL consistently achieves the highest correct selection rate across all settings, BIC performs slightly worse but remains competitive; compared to ICL, it tends to overestimate $K$. AIC consistently over-selects the number of components in all scenarios, showing a clear bias toward more complex models. Among the slope heuristic–based methods, Djump performs reasonably well, though still lags behind BIC and ICL in accuracy. In contrast, DDSE performs poorly across all settings. A potential explanation is that the relatively small value of $K_{\max}$ leads to underestimation of the penalty constant in the data-driven calibration procedure.

\begin{table}[htbp]
	\centering
	\renewcommand{\arraystretch}{1}
	\caption{Frequencies of selected component number $K$ by various criteria.}
	\begin{tabular}{c|lrrrrrrrrrrrrrr}\hline\hline
		$K_{\text{true}}$ & Method & 2 & 3 & 4 & 5 & 6 & 7 & 8 & 9 & 10 & 11 & 12 & 13 & 14 & 15 \\
		\hline
		\multirow{5}[0]{*}{2} 
		& AIC   &\textbf{1}  &6 &11  &9 & 9 &20 &11 &10 &11 & 6  &3 & 2  &1 & 0  \\
		& BIC   &\textbf{65} &13 &14  &5  &2  &1 & 0 & 0  &0  &0  &0 & 0  &0  &0 \\
		& ICL   &\textbf{73}   & 13   & 12     &0    & 1    & 1     &0   &  0     &0    & 0    & 0     &0   &  0    & 0 \\
		& DDSE  &\textbf{21} & 4  &4  &4  &4  &3  &2  &4  &2  &4  &6  &3 &10 &29  \\
		& Djump &\textbf{61}  &8 &13  &7  &3  &4  &2  &2  &0  &0  &0  &0  &0  &0  \\
		\hline
		\multirow{5}[0]{*}{3} 
		& AIC   &0  &\textbf{3}  &8  &6 &13 &15 &13 &20  &9  &2  &3  &4  &4  &0\\
		& BIC   &0 &\textbf{66} &15  &6 &10  &3  &0  &0  &0  &0  &0  &0  &0  &0  \\
		& ICL    &0    &\textbf{83}    &10     &4     &2    & 0    & 0     &1     &0    & 0     &0     &0    & 0  &   0 \\
		& DDSE  &0 &\textbf{17}  &2  &2  &4  &4  &3  &8  &4  &3  &3 &11 &17 &22 \\
		& Djump &2 &\textbf{66}  &9  &6 &11  &3  &1  &2 &0  &0  &0  &0  &0  &0\\
		\hline
		\multirow{5}[0]{*}{4} 
		& AIC   &0  &0  &\textbf{2}  &6 &12 &13 &16 &15 &16  &9  &8  &3  &0  &0  \\
		& BIC   &0  &0 &\textbf{73} &15  &8  &1  &3  &0  &0  &0  &0  &0  &0  &0 \\
		& ICL   &1     &2    &\textbf{78}    &11     &6    & 2     &0   &  0    & 0    & 0    & 0   &  0    & 0    & 0\\
		& DDSE  &0  &0 &\textbf{11}  &3  &3  &6  &6  &5  &8  &7  &7  &7 &12 &25 \\
		& Djump &2  &2 &\textbf{60} &15  &9  &8  &1  &2  &1  &0  &0  &0  &0  &0 \\
		\hline
		\multirow{5}[0]{*}{5} 
		& AIC   &0  &0  &0  &\textbf{6}  &6  &5 &16 &14 &16 &16  &6  &7  &6  &2  \\
		& BIC   &0  &0  &3 &\textbf{68} &14  &7  &4  &3  &1  &0  &0  &0  &0  &0\\
		& ICL   &9     &0     &2    &\textbf{70}    &11    & 3    & 2     &2     &0    & 0    & 0     &0     &1    & 0\\
		& DDSE  &0  &0  &0 &\textbf{11}  &3  &4  &4  &7  &9 &13  &4  &8 &12 &25 \\
		& Djump &2  &3  &9 &\textbf{56}  &9  &8  &5 & 4  &2  &2  &0  &0  &0 & 0  \\
		\hline\hline
	\end{tabular}%
	\label{tab:select_K_freq}
\end{table}

\FloatBarrier

\subsection{Parameter Estimation under Unknown Innovation Distributions with Data-Driven $K$}
Since the true innovation distribution is  unknown in practice, we evaluate how well the DAR model parameters are estimated using NM-QMLE when $K$ is selected using data-driven criteria (AIC, BIC, ICL, DDSE, and Djump) under various non-normal innovation scenarios, including $t$, skew-normal, and skew-$t$ distributions. For each scenario, we set the maximum number of components $K_{\max}=15$. The DAR(2) model parameters are set as $\theta_{1} = (\phi_{1}, \phi_2, \omega_{0}, \alpha_{1}, \alpha_2)^{T}=(0.3, 0.1, 1.0, 0.5, 0.2)^{T}$. For each simulation, we estimate the DAR(2) model for $K=2,\dots,15$ mixture components and apply the model selection criteria to choose the optimal number of components $\hat{K}$. Then we evaluate how the choice of $\hat{K}$ determined by each selection criterion influences the accuracy of the NM-QMLE parameter estimates for the DAR(2) model. The experiment is repeated 100 times. Estimation accuracy is measured by RMSE and mean bias.

\begin{table}[htbp]
	\centering
	\renewcommand{\arraystretch}{1.2} %
	\setlength{\tabcolsep}{5.2pt}     %
	\caption{RMSE and mean bias of estimated parameters for DAR(2) using NM-QMLE under different $K$ selection methods.}
	\resizebox{\textwidth}{!}{%
		\begin{tabular}{cl|cc|cc|cc|cc|cc}
			\hline\hline
			\multicolumn{2}{c|}{} 
			& \multicolumn{2}{c|}{AIC} 
			& \multicolumn{2}{c|}{BIC} 
			& \multicolumn{2}{c|}{ICL} 
			& \multicolumn{2}{c|}{DDSE} 
			& \multicolumn{2}{c}{Djump} \\
			\cline{3-12}
			\multicolumn{2}{c|}{} 
			& RMSE & Mean 
			& RMSE & Mean 
			& RMSE & Mean 
			& RMSE & Mean 
			& RMSE & Mean \\
			\hline
			\multirow{5}{*}{$t_{2.5}$} 
			& $\phi_1$   & 0.0488 & -0.0001 & 0.0353 &-0.0025  & 0.0377 &-0.0036  &0.0466   &-0.0023   &0.0330      &-0.0014     \\
			& $\phi_2$   & 0.0364 &0.0008       & 0.0304 &0.0004& 0.0308 &0.0024   &0.0377  &-0.0017     &0.0287      &0.0021       \\
			& $\omega$   & 0.3041 &-0.2201      & 0.5928 &-0.1740   & 0.2993 &-0.2683  &0.2952    &-0.2188  &0.2960   &-0.2311     \\
			& $\alpha_1$ & 0.2208 &-0.0747   & 0.3055 &-0.0896   & 0.1965 &-0.1244  &0.2100  &-0.0978    & 0.1867 &-0.1021     \\
			& $\alpha_2$ & 0.1059 &-0.0465   & 0.1861 &-0.0270 & 0.1036 &-0.0589  & 0.1081     &-0.0426      &0.1056     &-0.0507    \\
			\hline
			\multirow{5}{*}{skew normal(10)} 
			& $\phi_1$   &0.0304       &0.0019       &0.0248      &0.0005  &0.0275     &-0.0045   &0.0348   &-0.0001   &0.0274     &-0.0012       \\
			& $\phi_2$   &0.0327      &-0.0026    &0.0243      &-0.0008   &0.0262      &0.0043 &0.0370       &-0.0014      &0.0260      & -0.0008    \\
			& $\omega$   &0.1011    &0.0117     &0.0773      &0.0058  &0.0830      &0.0050 &0.0910     &0.0107    &0.0808     &0.0045      \\
			& $\alpha_1$ &0.0533    &0.0101     &0.0477    &0.0109  &0.0513     &0.0022  &0.0537   &0.0036   &0.0476      &0.0083      \\
			& $\alpha_2$ &0.0433 &-0.0014     &0.0302     &-0.0052 &0.0339    &-0.0010  &0.0407   &0.0032 &0.0331       &-0.0033    \\
			\hline
			\multirow{5}{*}{skewed $t(2.5, -0.9)$} 
			& $\phi_1$   &0.0362  &-0.0031        &0.0266     &-0.0016  &0.0294      &-0.0005  &0.0347     &-0.0058      &0.0284     &-0.0006   \\
			& $\phi_2$   &0.0328    & 0.0001    &0.0287   &-0.0046   &0.0279     &-0.0025   & 0.0331   &0.0001     &0.0274      &-0.0034      \\
			& $\omega$   &0.3837    &-0.1733       &0.4631      &-0.1785  &0.3294    &-0.2924 &0.3631     & -0.1927    & 0.3570    &-0.2030     \\
			& $\alpha_1$ &0.2028   &-0.1023        & 0.2341     &-0.1003  &0.1917    &-0.1404  &0.2005     & -0.1185 &0.2167       &-0.1177      \\
			& $\alpha_2$ &0.1512  &-0.0273    &0.1199      &-0.0412 &0.1023     &-0.0724  &0.1516     &-0.0333    &0.1447     &-0.0407   \\
			\hline\hline
		\end{tabular}%
	}
	\label{tab:selection_rmse_bias_extended}
\end{table}

Table \ref{tab:selection_rmse_bias_extended} shows that BIC, ICL, and Djump perform well in terms of their impact on DAR model parameter estimation, yielding accurate estimates with relatively low RMSE and mean bias. Compared to the results in Table \ref{tab:3}, where the number of components is fixed at $K=2$, these data-driven criteria provide only slight improvements in estimation accuracy. This suggests that fixing a small $K$ offers significant computational advantages while being sufficient for accurate parameter estimation in most cases.

\section{Empirical Study}
In the empirical study, we apply the newly proposed NM-QMLE to estimate the DAR model for the daily returns of the S\&P 500 index over the period from January 3, 2007, to December 29, 2023. Since the true underlying model is not directly observable, the performance of NMQMLE is evaluated through an application of testing the accuracy of Value at Risk~(VaR) estimate. 

VaR is a widely used method for measuring risk exposure in financial markets. It quantifies the potential loss of an asset or portfolio at a specified confidence level over a defined time horizon, such as a single day. A critical aspect of estimating VaR is the accurate modeling of the asset or portfolio's return distribution. In this context, the DAR model is fitted to the return distribution, capturing the dynamics of returns over time. Since VaR corresponds to the lower quantiles of the return distribution, an accurate estimation of the return model is essential for precise VaR calculations. More specifically, a better fit of the return distribution model leads to a more accurate estimate of these lower quantiles, directly improving the VaR estimate. Thus, the effectiveness of NM-QMLE in accurately estimating the model parameters is validated by its ability to provide more reliable VaR estimates.

For a DAR model, suppose $\{y_{t}, t=1, \cdots N \}$ is a series of return for a portfolio. Then VaR  of the portfolio at time $t$ for a given probability $p$ at time $t$, denoted as $q_{t} = q_{t}(p)$, is defined as
\begin{eqnarray*}
	q_{t} = \inf \{ x: p \leq \mathbb{P}_{t-1}(y_{t} \leq x) \} = \inf \{ x:  F_{t-1}(x) \geq p) \},
\end{eqnarray*}
where $\mathbb{P}_{t-1}$ and $F_{t-1}$ are the conditional probability and conditional cummulative distribution function of $y_{t}$ given all the information upto time $t-1$, respectively. From the DAR model specification $\eqref{Mod:DAR}$, one can show
\begin{eqnarray*}
	q_{t} = F^{-1}(p) = \phi_{0} {y}_{t-1} + \sqrt{h_{t}(\theta_{0})} F_{*}^{-1}(p),
\end{eqnarray*}
where $F_{*}^{-1}(p)$ is the $p^{th}$ quantile of the innovation $\{\eta_{t}\}$ conditional on all the information available upto time $t-1$, and $h_t(\theta_0)=\omega_0+\alpha_0 y_{t-1}^2$ is the conditional variance of the return. When substituting estimated parameters for the true ones, we replace $h_{t}(\theta_{0})$ with the estimator $h_{t}(\widehat{\theta}_{t})$, and the quantile $F_{*}^{-1}(p)$ is estimated by the the $p^{th}$ quantile of the standardized residuals $({y}_{t} - \phi {y}_{t-1})/h_{t}^{1/2}(\widehat{\theta}_{t})$. The newly proposed NM-QMLE for the DAR model is applied to estimate the parameters using the estimation sample, which consists of all observations from $T_0$ (the starting point for estimation) to $T_1 = t-1$. The estimated parameters are then used to calculate the VaR $ \widehat{q}_{t}$ for day $t$ in the testing sample.

\begin{eqnarray*}
	\widehat{q}_{t} = \widehat{\phi} {y}_{t-1} + \sqrt{h_{t}(\widehat{\theta}_{t})}
	\displaystyle\bigg(
	({y}_{t} - \widehat{\phi} {y}_{t-1})/h_{t}^{1/2}(\widehat{\theta}_{t}),\  T_0 \leq t \leq T_1
	\displaystyle\bigg)_{p}, \quad  1 \leq t \leq N,
\end{eqnarray*}
where $(X_{t})_{p}, T_0 \leq t \leq T_1$ denotes the $(\lceil p*n \rceil)^{th}$ ordered statistics of  $\{ X_{t}, T_0 \leq t \leq T_1 \}$, with $\lceil x \rceil$ representing the ceiling function.

We analyze the performance of the NMQMLE using real financial data by estimating the VaR.  To assess the accuracy of the model, we apply the likelihood ratio tests proposed in \cite{K1995} and \cite{C1998}. Define
\begin{eqnarray*}
	N_{*} = \sum_{t=1}^{N} I_{t},  I_{t}=I(y_{t} \leq \hat{q}_{t}), \quad
	\hat{p} = N_{*}/N.
\end{eqnarray*}

According to \cite{K1995}, the likelihood ratio statistic for the proportion of failures (POF) test is defined as
\begin{eqnarray*}
	LR_{POF} = 2 \displaystyle\bigg[
	\log \{(1-\hat{p})^{N-N_{*}} \hat{p}^{N_{*}}\} - \log \{(1-p)^{N-N_{*}} p^{N_{*}}\}
	\displaystyle\bigg].
\end{eqnarray*}

Let $N_{i,j}$ denote the number of observations in $\{ t: I_{t-1}=j, I_{t}=i, 2 \leq t \leq N \}$, where $i, j \in \{0,1\}$. The transition probabilities and the unconditional failure rate are defined as
\begin{eqnarray*}
	\hat{\tau}_{i,j} = \frac{N_{i,j}}{(N_{i,0}+N_{i,1})}, \text{ and }  \hat{\tau} = \frac{(N_{0, 1}+N_{1, 1})}{N}.
\end{eqnarray*}

Following \cite{C1998}, the conditional coverage independence (CCI) test statistic is given by
\begin{eqnarray*}
	LR_{CCI} = 2 \displaystyle\bigg[
	\log \{(1-\hat{\tau}_{0,1})^{N_{0,0}} \hat{\tau}_{0,1}^{N_{0,1}} (1-\hat{\tau}_{1,1})^{N_{1,0}}
	\hat{\tau}_{1,1}^{N_{1,1}} \} -
	\log \{(1-\hat{\tau})^{N_{0, 0}+N_{1, 0}} \hat{\tau}^{N_{0, 1}+N_{1, 1}}\}
	\displaystyle\bigg],
\end{eqnarray*}
The combined likelihood ratio statistic for conditional coverage (CC) is then
\begin{eqnarray*}
	LR_{CC} = LR_{POF} + LR_{CCI},
\end{eqnarray*}
According to \cite{C1998}, the asymptotic distribution of the test statistic $LR_{CC}$ follows $\chi_{(2)}^{2}$ distribution.

\begin{figure}[htbp]
	\centering
	\includegraphics[width=0.7\textwidth]{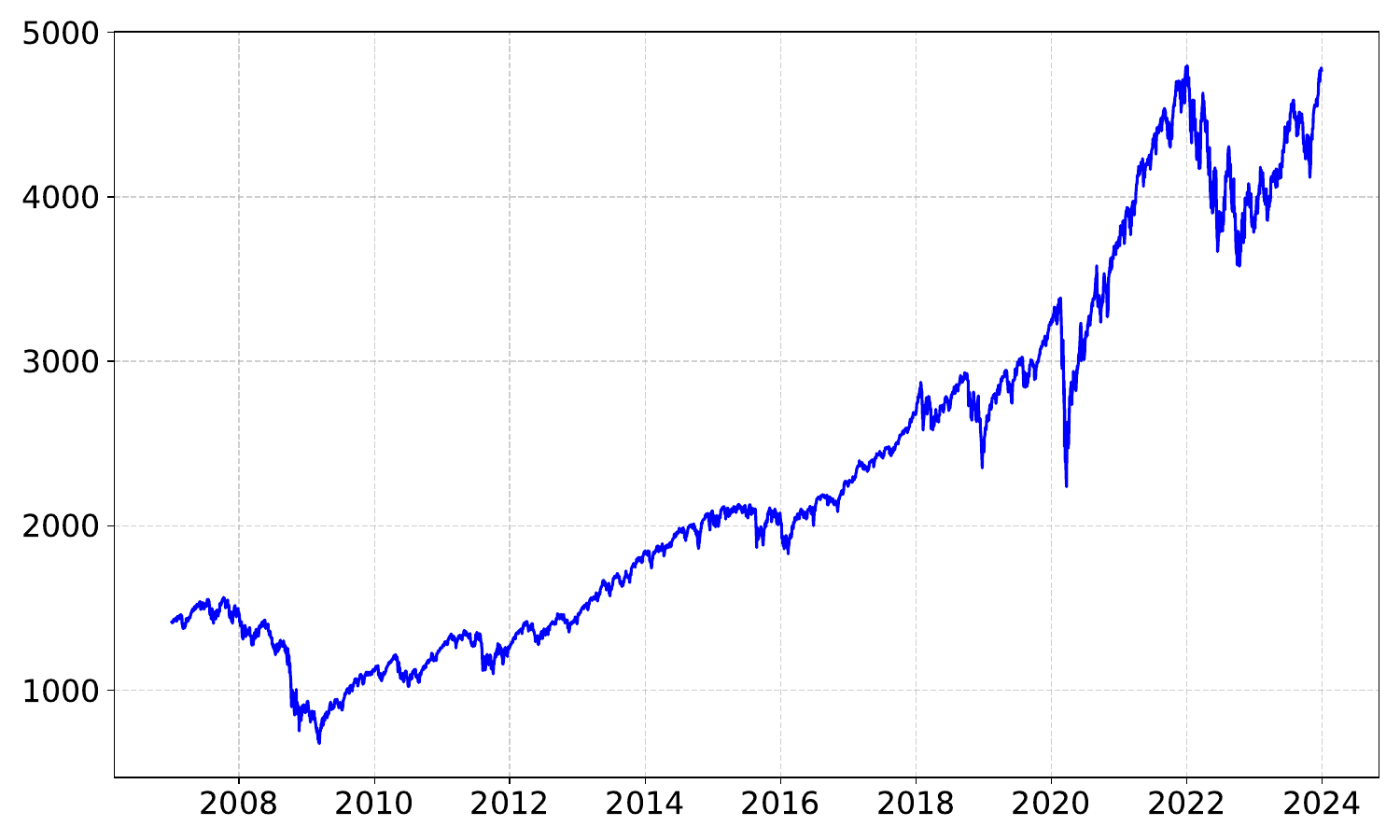}
	\caption{The daily close price of S\&P 500 index from  January 3, 2007, to December 29, 2023.}  \label{fig42}
\end{figure}

Figure \ref{fig42} illustrates the daily closing prices of the S\&P 500 index from January 3, 2007, to December 29, 2023, a period that encompasses two extreme market events: the 2008 financial crisis and the 2020 COVID-19 pandemic. The parameters of the DAR(1) model are estimated using NMQMLE based on the returns. To compute $\widehat{q}_{t}$, we use the period from January 3, 2005, to December 29, 2006, as the initial estimation window for calculating the VaR on January 3, 2007. Subsequently, starting from January 3, 2007, to December 29, 2023, the estimation window is expanded daily by including one additional observation, and the parameters are re-estimated to compute $\widehat{q}_{t}$ for each day. For different significance levels, $p= 0.01, 0.025, 0.05$,  the estimated $\widehat{q}_{t}$ values and the test statistics $LR_{CC}$ are calculated using the fitted DAR model. The testing results are summarized in Table \ref{tableAPPlication4}.

\begin{table}[htbp]
	\centering
	\caption{\label{tableAPPlication4} Testing results according to $LR_{CC}$ for VaR of S\&P 500 index.}
	\begin{tabular}{c |c}
		\hline
		\hline
		Significant level $p$ & $LR_{CC}$ \\
		\hline
		{$1\%$}   &  2.8779   \\
		\hline
		{$2.5\%$} &  1.1553  \\
		\hline
		{$5\%$}   &  3.8709  \\
		\hline
		\hline
	\end{tabular}\\
\end{table}

By comparing the observed values with the critical values from the $\chi^2$ distribution: $\chi^{2}_{(2)}(0.01) = 9.21$, $\chi^{2}_{(2)}(0.025) = 7.38$,  $\chi^{2}_{(2)}(0.05) = 5.99$, we conclude that the null hypothesis cannot be rejected. In other words, we do not reject the accuracy of the estimated VaR derived from the DAR model parameters estimated using NM-QMLE for this dataset. This result suggests that the NM-QMLE provides sufficiently accurate estimations, even for the tail distribution.
\FloatBarrier

\section{Conclusion}
\label{sec:conc}
In this paper, we propose a new estimation method, the Normal Mixture Quasi-Maximum Likelihood Estimation (NM-QMLE), for Double Autoregressive (DAR) models, specifically to address scenarios where the innovation distribution is unknown or significantly non-normal. By substituting the unknown true distribution of innovations with a normal mixture density within a quasi-likelihood framework, our approach captures both heavy-tailed behavior and skewness in the data, leading to more reliable parameter estimates. We establish the consistency and asymptotic normality of the NM-QMLE estimator under regularity conditions. Through Monte Carlo simulations, we demonstrate that our method outperforms existing approaches, achieving lower estimation errors in finite samples. Furthermore, an empirical analysis using daily returns of the S\&P 500 index highlights the practical utility of our method in generating accurate VaR estimates for financial risk management.

\bibliographystyle{Chicago}

\newpage
\appendix
\section{Data generation of skew normal distribution}

The density function of a skew normal random variable $U$ with parameter $\theta$ is
\begin{eqnarray*}
	f(x) = 2 \varphi (x) \Phi (\theta x),
\end{eqnarray*}
where  $\varphi$ and $\Phi$ represent the density function and cumulative distribution function of standard normal distribution respectively.
Suppose $U^{1}$ and $U^{2}$ are independent and identically distributed (i.i.d.) normal random variables. Then the skew normal random variable can be generated by
\begin{eqnarray*}
	\frac{\theta |U^{1}| + U^{2}}{\sqrt{1 + \theta^{2}}}.
\end{eqnarray*}
Similarly, a sample $\{U_{i}, i = 1, \cdots, n\}$ from skew normal distribution can be generated by first simulating two independent standard normal samples:
\begin{eqnarray*}
	\{U^{1}_{i}, i = 1, \cdots, n\}, \quad \{U^{2}_{j}, j = 1, \cdots, n\},
\end{eqnarray*}
and then constructing $U_i$ by
\begin{eqnarray*}
	U_{i} = \frac{\theta |U^{1}_{i}| + U^{2}_{i}}{\sqrt{1 + \theta^{2}}}, \quad  i = 1, \cdots, n.
\end{eqnarray*}

\section{Data generation of skewed t distribution}

The density function of a skew t random variable $X$ with parameter $(\mu, \sigma, q,\ \lambda)$ is
\begin{eqnarray*}
	f(x) = \frac{\Gamma(\frac{1+q}{2})}{\nu\sigma(\pi q/2)^{1/2}\Gamma(\frac{q}{2})(\frac{|x-\mu+m|^2}{\frac{q}{2}(\nu\sigma)^2(\lambda sign(x-\mu+m)+1)^2})^{\frac{1+q}{2}}},
\end{eqnarray*}
In our study, without loss of generosity, we take $\mu = 0$ and $\sigma = 1$. $q$ controls the kurtosis and $\lambda$
controls the skewness. Skewed t distributions are generated according to the above formula with
different parameterizations of $q$ and $\lambda$.

\section{Lemmas}

We will prove Theorem \ref{theorem41} and Theorem \ref{theorem42} by the approach in \cite{A1985}. There are a few assumptions to be verified first and we need the following lemmas.

\begin{lem}
	\label{lem41}
	Suppose Assumptions~(A1 - A4) hold, then it follows that
	\begin{eqnarray*}
		\E \displaystyle\bigg\{\sup_{\btheta \in \bTheta }|W_{t}(\btheta)| \displaystyle\bigg\} <\infty,
	\end{eqnarray*}
	and
	\begin{eqnarray*}
		\sup_{\btheta \in \bTheta } \abs{l_{n}(\btheta)- \E W_{t}(\btheta)} = o_{p}(1), \quad \mbox{as } n \rightarrow \infty.
	\end{eqnarray*}
\end{lem}
\vspace{0.1in}

\noindent\textbf{\underline{Proof}:}
We use the similar strategy in Lemma 1 of \cite{L2004}. Recall $\btheta_1 = (\bphi^T, \omega, \balpha^T)^T$ and $h_t(\omega, \balpha) = \omega + \alpha_1 y^{2}_{t-1} + \ldots + \alpha_p y^{2}_{t-p}$. We have
\beqn\label{equ1}
&  & \E
\displaystyle\bigg\{
\sup_{\btheta_{1} \in \bTheta_{1} } \abs{ \log (\omega + \alpha_1 y^{2}_{t-1} + \ldots + \alpha_p y^{2}_{t-p})}
\displaystyle\bigg\} \nonumber \\
&\leq  & \E \displaystyle\bigg\{
\sup_{\btheta_{1} \in \bTheta_{1} } \log h_t(\omega, \balpha) \cdot
I(h_t(\omega, \balpha) \geq 1)\displaystyle\bigg\} + \E \displaystyle\bigg\{
\sup_{\btheta_{1} \in \bTheta_{1} } \log h_t(\omega, \balpha) \cdot[- I(h_t(\omega, \balpha) < 1)] \displaystyle\bigg\} \nonumber \\
&\leq  & \E \log \left[ 1+ \bar\omega + \bar\alpha(y^{2}_{t-1} + \ldots + y^{2}_{t-p}) )\right] - \log\underline\omega \cdot
I(\underline\omega <1),
\eeqn
where the last inequality holds since the condition~(A2). For the first term in (\ref{equ1}), we have
\beqn\label{equ2}
&  &  \E \log \left[ 1+ \bar\omega + \bar\alpha(y^{2}_{t-1} + \ldots + y^{2}_{t-p}) )\right] \nonumber \\
&\leq&  \log \left[ (1+ \bar\omega) + \bar\alpha
(\E\abs{y}^{2}_{t-1} + \ldots + \E\abs{y}^{2}_{t-p}) \right]
\eeqn
The equation~\eqref{equ2} is obtained by Jensen's inequality. By assumption that $\E\abs{y_t}^2 < \infty$ for any $t$, it follows that
\beq
\E \sup_{\btheta_{1} \in \bTheta_{1} } \abs{\log h_t(\omega, \balpha)} < \infty.
\eeq
According to Definition (\ref{WtDef}),
\begin{eqnarray*}
	W_{t}(\btheta)
	&=&
	\log
	\displaystyle\bigg\{
	\frac{1}{\sqrt{h_t(\omega, \balpha)}}
	g_{\btheta_{2}}
	\displaystyle\bigg(\frac{ y_{t} - m_{t}(\bphi)}{\sqrt{h_t(\omega, \balpha)}} \displaystyle\bigg)
	\displaystyle\bigg\}   \\
	&=&
	\log \displaystyle\bigg\{
	\frac{1}{\sqrt{h_t(\omega, \balpha)}}
	\sum ^{K}_{k=1}  p_{k} \frac{1}{\sqrt{2\pi}\sigma_{k}}
	\exp
	\displaystyle\bigg[
	- \frac{1}{2\sigma^{2}_{k}} \left(\frac{ y_{t} - m_{t}(\bphi)}{\sqrt{h_t(\omega, \balpha)}} - \mu_{k} \right)^{2}
	\displaystyle\bigg]
	\displaystyle\bigg\}.
\end{eqnarray*}
Since $e^x < 1,~ x<0$, we obtain that
\begin{align*}
	\E \sup_{\btheta \in \bTheta } \abs{ W_{t}(\btheta) }
	&=~
	\E \sup_{\btheta \in \bTheta }
	\displaystyle\bigg |
	\log \displaystyle\bigg\{
	\frac{1}{\sqrt{h_t(\omega, \balpha)}}   \\
	& \quad \quad \sum ^{K}_{k=1}  p_{k} \frac{1}{\sqrt{2\pi}\sigma_{k}}
	\exp
	\displaystyle\bigg[
	- \frac{1}{2\sigma^{2}_{k}} \left(\frac{ y_{t} - m_{t}(\bphi)}{\sqrt{h_t(\omega, \balpha)}} - \mu_{k} \right)^{2}
	\displaystyle\bigg]
	\displaystyle\bigg\}
	\displaystyle\bigg |
	\\
	& \leq~
	\frac{1}{2}
	\E \sup_{\btheta_1 \in \bTheta_1 } \abs{ \log h_t(\omega, \balpha) } +
	\E \sup_{\btheta_2 \in \bTheta_2 } \abs{
		\log \displaystyle\bigg\{
		\sum ^{K}_{k=1}  \frac{ p_{k} }{\sqrt{2\pi}\sigma_{k}}
		\displaystyle\bigg\} }
	\\
	&< \infty.
\end{align*}
From Assumption ~(A3), $\{ y_{t}, t = 0,1, \cdots\}$ is a strictly stationary and ergodic time series.
Next we will show that
\begin{eqnarray*}
	\sup_{\btheta \in \bTheta } \abs{ \frac{1}{n} \sum^{n}_{t=1} W_{t}(\btheta) -\E W_{t}(\btheta) }= o_{p}(1).
\end{eqnarray*}
Since $\bTheta$ is a compact set, we can partition $\bTheta$ into $m$ non-overlapping region
$\bTheta_1^m, \ldots, \bTheta_m^m$. Let $\btheta_1, \ldots, \btheta_m$ be a sequence of vectors
such that $\btheta_i \in \bTheta_i^m$. It follows that
\beqn\label{equ8.4}
&  & \Pr \left( \sup_{\btheta \in \bTheta} \abs{ \frac1n \sum_{t=1}^n (W_t(\btheta) - \E W_t(\btheta)) } > \eps   \right) \nonumber \\
&\leq& \Pr \left( \bigcup_{i=1}^m \left\{ \sup_{\btheta \in \bTheta_i^m} \abs{ \frac1n \sum_{t=1}^n (W_t(\btheta) - \E W_t(\btheta)) } > \eps  \right\} \right) \nonumber \\
&\leq& \sum_{i=1}^m \Pr \left( \sup_{\btheta \in \bTheta_i^m} \abs{ \frac1n \sum_{t=1}^n (W_t(\btheta) - \E W_t(\btheta)) } > \eps  \right) \nonumber \\
&\leq& \sum_{i=1}^m \Pr \left( \sup_{\btheta \in \bTheta_i^m} \abs{ \frac1n \sum_{t=1}^n (W_t(\btheta_i) - \E W_t(\btheta_i)) } > \frac{\eps}{2}  \right) \nonumber \\
& & \quad + \sum_{i=1}^m \Pr \left( \sup_{\btheta \in \bTheta_i^m} \abs{ \frac1n \sum_{t=1}^n (W_t(\btheta) -  W_t(\btheta_i) - \E W_t(\btheta) + \E W_t(\btheta_i) ) } > \frac{\eps}{2}  \right)
\eeqn
Because $W_t(\btheta)$ and $\E W_t(\btheta)$ is uniformly continuous in $\btheta \in \bTheta$, we have that, for
each $i = 1, \ldots, m$,
\begin{align}
	&\lim_{m \to \infty} \sup_{\btheta \in \bTheta_i^m} \abs{ W_t(\btheta) - W_t(\btheta_i)} = 0; \\
	&\lim_{m \to \infty} \sup_{\btheta \in \bTheta_i^m} \abs{ \E W_t(\btheta) -  \E W_t(\btheta_i)} = 0.
\end{align}
With the facts
\beqn
& & \sup_{\btheta \in \bTheta_i^m} \abs{W_t(\btheta) - W_t(\btheta_i)}  \nonumber \\
&\leq& \sup_{\btheta \in \bTheta} \abs{W_t(\btheta) } + \sup_{\btheta_i \in \bTheta} \abs{W_t(\btheta_i) } \nonumber \\
& = & 2 \sup_{\btheta \in \bTheta} \abs{W_t(\btheta) }
\eeqn
and $\E \sup_{\btheta \in \bTheta} \abs{W_t(\btheta) } < \infty$, by DCT, it follows that
\beq
\lim_{m \to \infty} \E \sup_{\btheta \in \bTheta_i^m} \abs{ W_t(\btheta) -  W_t(\btheta_i) }
= \E \lim_{m \to \infty} \sup_{\btheta \in \bTheta_i^m} \abs{ W_t(\btheta) -  W_t(\btheta_i) } = 0,
\eeq
uniformly for each $i$. This means that $\E \sup_{\theta \in \Theta_i^m} \abs{ W_t(\btheta) -  W_t(\btheta_i) } < \eps/2$ for any $\eps>0$, when $m$ is sufficiently large. Finally, by the ergodic theorem, the two terms in \eqref{equ8.4} are both zeros. We obtain the lemma 3.

\begin{lem}
	\label{lem42}
	Given Assumptions~(A1-A5), $\E W_{t}(\btheta)$ attains a strict local maximum at $\btheta=\btheta_{0}$.
\end{lem}

\noindent\textbf{\underline{Proof:}}
From Assumption (A4),
for any parameter $\btheta_{2}(\neq \btheta_{20}) \in \bTheta_{2}, u(\neq 1) > 0$, there is
\begin{eqnarray*}
	\E \log \{ u g_{\btheta_{2}}(u \eta_{t}) \} < \E \log g_{\btheta_{20}}(\eta_t).
\end{eqnarray*}
Then for any $\btheta \in \bTheta$, we have
\begin{align*}
	&\quad~ \E W_{t}(\btheta) - \E W_{t}(\btheta_{0})  \\
	&= \E \left(
	\E \left( W_{t}(\btheta) | \eps_{t-1}, \eps_{t-2}, \cdots \right)
	\right)
	- \E \left(
	\E \left( W_{t}(\btheta_{0}) | \eps_{t-1}, \eps_{t-2}, \cdots \right)
	\right)
	\\
	&=
	\E \left\{
	\E \left(
	\log  \left\{
	\frac{1}{\sqrt{h_t(\omega, \balpha)}}
	g_{\btheta_{2}}
	\left(\frac{ y_{t} - m_{t}(\bphi)}{\sqrt{h_t(\omega, \balpha)}} \right)
	\right\}  \bigg | \eps_{t-1}, \eps_{t-2}, \cdots \right) \right\}  \\
	& \quad -
	\E \left\{
	\E \left(
	\log  \left\{
	\frac{1}{\sqrt{h_t(\omega_0, \balpha_0)}}
	g_{\btheta_{20}}
	\left( \frac{ y_{t} - m_{t}(\bphi_0)}{\sqrt{h_t(\omega_0, \balpha_0)}} \right)
	\right\}
	\displaystyle\bigg | \eps_{t-1}, \eps_{t-2}, \cdots \right)
	\right\}  \\
	&=
	\E \displaystyle\bigg(
	\log \frac{1}{\sqrt{h_t(\omega, \balpha)}}  \\
	& \quad \quad + \E \log g_{\btheta_{2}} \left(
	\frac{\sqrt{h_t(\omega_{0}, \balpha_{0})}}{\sqrt{h_t(\omega, \balpha)}} \eta_{t}
	+ \frac{m_t(\bphi_{0}) - m_t(\bphi)}{\sqrt{h_t(\omega, \balpha)}} \right)
	\displaystyle\bigg | \eps_{t-1}, \eps_{t-2}, \cdots \displaystyle\bigg)  \\
	& \quad \quad -
	\E \left\{ \E \left(
	\log  \left\{
	\frac{1}{\sqrt{h_t(\omega_{0}, \balpha_{0})}}
	g_{\btheta_{20}} \left(
	\frac{ y_{t} - m_t(\bphi_{0})}{\sqrt{h_t(\omega_{0}, \balpha_{0})}} \right)
	\right\}
	\displaystyle\bigg | \eps_{t-1}, \eps_{t-2}, \cdots \right) \right\}  \\
	&\leq
	\E \left\{
	\E \left(
	\log  \left\{
	\frac{1}{\sqrt{h_t(\omega, \balpha)}}
	g_{\btheta_{2}} \left(
	\frac{\sqrt{h_t(\omega_{0}, \balpha_{0})}}{\sqrt{h_t(\omega, \balpha)}} \eta_{t}
	\right) \right\}
	\displaystyle\bigg | \eps_{t-1}, \eps_{t-2}, \cdots \right) \right\} \\
	& \quad \quad -
	\E \left\{ \E \left(
	\log  \left\{
	\frac{1}{\sqrt{h_t(\omega_{0}, \balpha_{0})}}
	g_{\btheta_{20}} \left(
	\frac{ y_{t} - m_t(\bphi_{0})}{\sqrt{h_t(\omega_{0}, \balpha_{0})}} \right)
	\right\}
	\displaystyle\bigg | \eps_{t-1}, \eps_{t-2}, \cdots \right) \right\}  \\
	&= \E
	\displaystyle\bigg\{
	\E \displaystyle\bigg(
	\log \left\{
	\frac{\sqrt{h_t(\omega_{0}, \balpha_{0})}}{\sqrt{h_t(\omega, \balpha)}}
	g_{\btheta_{20}}
	\left(\frac{\sqrt{h_t(\omega_{0}, \balpha_{0})}}{\sqrt{h_t(\omega, \balpha)}} \eta_{t}
	\right)
	\right\}
	- \log g_{\btheta_{20}}(\eta_{t})
	\displaystyle\bigg | \eps_{t-1}, \eps_{t-2}, \cdots
	\displaystyle\bigg\}
	\displaystyle\bigg\}  \\
	&\leq
	\E \displaystyle\bigg\{
	\E \displaystyle\bigg\{
	\log g_{\btheta_{20}}(\eta_{t})
	-
	\log g_{\btheta_{20}}(\eta_{t})
	\displaystyle\bigg | \eps_{t-1}, \eps_{t-2}, \cdots
	\displaystyle\bigg\}
	\displaystyle\bigg\}  \\
	& = 0,
\end{align*}
where the penultimate equation holds if and only if $\bphi=\bphi_{0}$, $h_t(\omega, \balpha) =
h_t(\omega_{0}, \balpha_{0})$, $\btheta_{2}=\btheta_{20}$ that is $\btheta=\btheta_{0}$.
Then Lemma \ref{lem42} is obtained

\begin{lem}
	\label{lem43}
	Suppose that Assumptions ~(A1-A6) hold, it follows that
	\begin{eqnarray}
		\E
		\sup_{\btheta \in \bTheta } \left\|
		\frac{\partial W_{t}(\btheta)}{\partial \btheta}\frac{\partial W_{t}(\btheta)}{\partial \btheta^{T} }\right\|
		< \infty, \label{lem43-1}
	\end{eqnarray}
	and
	\begin{eqnarray}
		\sup_{\btheta \in \bTheta }
		\left\| \frac{1}{n} \sum ^{n}_{t=p+1}
			\left(  \frac{\partial W_{t}(\btheta)}{\partial \btheta}\frac{\partial W_{t}(\btheta)}{\partial \btheta^{T} } - \E \frac{\partial W_{t}(\btheta)}{\partial \btheta}\frac{\partial W_{t}(\btheta)}{\partial \btheta^{T} }
			\right) \right\|
		= o_{p}(1), \quad n \rightarrow \infty.\label{lem43-2}
	\end{eqnarray}
\end{lem}

\vspace{0.1in}\

\noindent\textbf{\underline{Proof:}} \ Denote by $\widetilde y_t(\btheta_1) = (y_{t} - m_{t}(\bphi))
/\sqrt{h_t(\omega, \balpha)}$ and
\beqnn
f_{k} = \frac{1}{\sqrt{2\pi}\sigma_{k}}
\exp \displaystyle\bigg\{
- \frac{1}{2\sigma^{2}_{k}} \left(\frac{y_{t} - m_t(\bphi)}{h_t(\omega, \balpha)} - \mu_{k}\right)^{2}
\displaystyle\bigg\}
= \frac{1}{\sqrt{2\pi} \sigma_{k}}
\exp \displaystyle\bigg\{ -\frac{(\widetilde y_t(\btheta_1) - \mu_{k})^{2}}{2 \sigma_{k}^{2}} \displaystyle\bigg\}.
\eeqnn
Combined with the definition (\ref{WtDef}) of $W_{t}(\btheta)$, we have
\begin{align}
	W_{t}(\btheta)
	&=
	\log
	\left(
	\frac{1}{\sqrt{h_t (\omega, \balpha) }}
	g_{\btheta_{2}} (\frac{ y_{t} - m_t(\bphi)}{h_t(\omega, \balpha)} )
	\right)   \nonumber  \\
	&=
	\log \left(
	\frac{1}{\sqrt{h_t(\omega, \balpha)}}
	\sum ^{K}_{k=1}  p_{k}
	\frac{1}{\sqrt{2\pi}\sigma_{k}}
	\exp \left\{
	- \frac{1}{2\sigma^{2}_{k}}(\frac{ y_{t} - m_t(\bphi)}{h_t(\omega, \balpha)} - \mu_{k})^{2}
	\right\}
	\right)   \nonumber   \\
	&=
	- \frac{1}{2} \log h_t(\omega, \balpha)
	+ \log \left( \sum ^{K}_{k=1}  p_{k} f_{k} \right).  \label{WtDef2}
\end{align}
Next, we calculate $\partial W_{t}(\btheta)/\partial \btheta$, where
\begin{align*}
	\theta
	&= (\btheta_{1}^T ,\btheta_{2}^T )^T   \\
	&= (\bphi^T, \omega, \balpha^T, p_{1}, \cdots, p_{k-1}, \mu_{1}, \cdots, \mu_{k-1}, \sigma_{1}, \cdots, \sigma_{k-1})^T.
\end{align*}
Based on equation (\ref{WtDef2}) of $W_{t}(\btheta)$, we obtain the derivatives,
for $j = 1, \ldots, p$, and $k < K$,
\begin{align}
	\frac{\partial W_{t}(\btheta)}{\partial \phi_j}
	&= \frac{1}{\sum ^{K}_{k=1}  p_{k} f_{k}}
	\sum ^{K}_{k=1} p_{k} f_{k} \frac{y_{t-j}(\widetilde y_t(\btheta_1) -\mu_{k})}{\sigma_{k}^{2} h_t^{\frac{1}{2}}(\omega, \balpha)}, \\
	\frac{\partial W_{t}(\btheta)}{\partial \omega}
	&= \frac{1}{\sum ^{K}_{k=1} p_{k} f_{k}}
	\sum ^{K}_{k=1} p_{k} f_{k}
	\frac{\widetilde y_t(\btheta_1)( \widetilde y_t(\btheta_1) -\mu_{k})}{2 \sigma_{k}^{2} h_t(\omega, \balpha)}
	- \frac{1}{2 h_t(\omega, \balpha)} , \\
	\frac{\partial W_{t}(\btheta)}{\partial \alpha_j}
	&= \frac{y_{t-j}^{2}}{\sum ^{K}_{k=1} p_{k} f_{k}}
	\sum ^{K}_{k=1} p_{k} f_{k}
	\frac{\widetilde y_t(\btheta_1)( \widetilde y_t(\btheta_1) -\mu_{k})}{2 \sigma_{k}^{2} h_t(\omega, \balpha)}
	- \frac{y_{t-j}^{2}}{2 h_t(\omega, \balpha)}, \\
	\frac{\partial W_{t}(\btheta)}{\partial p_{k}}
	&= \frac{1}{\sum ^{K}_{k=1} p_{k} f_{k}}
	\displaystyle\bigg(
	f_{k} + p_{K} \frac{\partial f_{K}}{\partial p_{k}} +f_{K} \frac{\partial p_{K}}{\partial p_{k}}
	\displaystyle\bigg), \\
	\frac{\partial W_{t}(\btheta)}{\partial \mu_{k}}
	&= \frac{1}{\sum ^{K}_{k=1} p_{k} f_{k}}
	\displaystyle\bigg(
	\sum_{k=1}^{K-1} p_k f_k \frac{\widetilde y_t(\btheta_1) - \mu_k}{\sigma_k^2} + p_K \frac{\partial f_K}{\partial \mu_{k}}
	\displaystyle\bigg), \\
	\frac{\partial W_{t}(\btheta)}{\partial \sigma_{k}}
	&= \frac{1}{\sum ^{K}_{k=1} p_{k} f_{k}}
	\displaystyle\bigg(
	\sum_{k=1}^{K-1} p_k f_k \big(-\frac{1}{\sigma_k} + \frac{\widetilde y_t(\btheta_1) - \mu_k}{\sigma_k^3}\big) + p_K \frac{\partial f_K}{\partial \sigma_{k}}
	\displaystyle\bigg).
\end{align}
Since the constraints on
$\btheta_2 = (p_{1},\ldots,p_{k-1},\mu_{1},\ldots,\mu_{k-1},\sigma_{1},\cdots,\sigma_{k-1})^T$ is that
\begin{align*}
	p_{K}
	&= 1 - \sum^{K-1}_{k=1} p_{k},   \\
	\mu_{K}
	&= \frac{1-\sum^{K-1}_{k=1} p_{k}\mu_{k}}{p_{K}},   \\
	\sigma_{K}
	&=
	\displaystyle\bigg(
	\frac{1-\sum^{K-1}_{k=1} p_{k}(\mu_{k}^{2} + \sigma_{k}^{2}) - p_{K}\mu^{2}_{K}}{p_{K}}
	\displaystyle\bigg) ^{\frac{1}{2}},
\end{align*}
we further have the derivatives
\begin{align*}
	\frac{\partial p_{K}}{\partial p_{k}}
	&= -1,   \\
	\frac{\partial \mu_{K}}{\partial p_{k}}
	&= \frac{\mu_{K} - \mu_{k}}{p_{K}},   \\
	\frac{\partial \sigma_{K}}{\partial p_{k}}
	&=
	\frac{\sigma_{K}^{2} - \mu_{K}^{2} + 2\mu_{k}\mu_{K} - (\mu_{k}^{2}+\sigma_{k}^{2})}{2p_{K}\sigma_{K}},   \\
	\frac{\partial f_{K}}{\partial p_{k}}
	&= f_{K}
	\displaystyle\bigg(
	-\frac{1}{\sigma_{K}}  \frac{\partial \sigma_{K}}{\partial p_{k}}
	+ \frac{\widetilde y_t(\btheta_1)-\mu_{K}}{\sigma_{K}^{2}} \frac{\partial \mu_{K}}{\partial p_{k}}
	+ \frac{(\widetilde y_t(\btheta_1)-\mu_{K})^{2}}{\sigma_{K}^{3}} \frac{\partial \sigma_{K}}{\partial p_{k}}
	\displaystyle\bigg); \\
	\\
	\frac{\partial \mu_{K}}{\partial \mu_{k}}
	&= -\frac{p_k}{p_K}, \\
	\frac{\partial \sigma_{K}}{\partial \mu_{k}},
	&= \frac{p_k(\mu_K - \mu_k)}{p_K \sigma_K},
	\\
	\frac{\partial f_{K}}{\partial \mu_{k}}
	&= f_K\displaystyle\bigg(
	-\frac{1}{\sigma_{K}}  \frac{\partial \sigma_{K}}{\partial \mu_{k}}
	+ \frac{\widetilde y_t(\btheta_1) -\mu_{K}}{\sigma_{K}^{2}} \frac{\partial \mu_{K}}{\partial \mu_{k}}
	\displaystyle\bigg);  \\
	\\
	\frac{\partial \sigma_{K}}{\partial \sigma_{k}}
	&= -\frac{p_k \sigma_k}{p_K \sigma_K},      \\
	\frac{\partial f_{K}}{\partial \sigma_{k}}
	&= f_K\displaystyle\bigg(
	-\frac{1}{\sigma_{K}} + \frac{(\widetilde y_t(\btheta_1) -\mu_{K})^2}{\sigma_{K}^{3}}
	\displaystyle\bigg)\frac{\partial \sigma_{K}}{\partial \sigma_{k}}.
\end{align*}
By Jensen's inequality, then there exists a constant $C>0$ such that
\begin{align*}
	\left( \frac{\partial W_{t}(\btheta)}{\partial \phi_j} \right)^2
	&=
	\displaystyle\bigg (
	\sum ^{K}_{k=1}
	\frac{p_{k} f_{k}} {\sum ^{K}_{k=1} p_{k} f_{k}}
	\frac{y_{t-j}(\widetilde y_t(\btheta_1) -\mu_{k})}{\sigma_{k}^{2}\sqrt{h_t(\omega, \balpha)}}
	\displaystyle\bigg )^{2}   \\
	&\leq
	\sum ^{K}_{k=1}
	\frac{p_{k} f_{k}} {\sum ^{K}_{k=1} p_{k} f_{k}}
	\left(\frac{y_{t-j}(\widetilde y_t(\btheta_1) -\mu_{k})}{\sigma_{k}^{2}\sqrt{h_t(\omega, \balpha)}}
	\right)^{2}   \\
	&\leq
	C \sum ^{K}_{k=1} \frac{y_{t-j}^{2}(\widetilde y_t(\btheta_1) -\mu_{k})^{2}}{h_t(\omega, \balpha)}   \\
	&\leq
	2C \sum ^{K}_{k=1} \frac{y_{t-j}^{2}(\widetilde y_t(\btheta_1)^2 +\mu_{k}^{2})}{h_t(\omega, \balpha)}.
\end{align*}
Furthermore, we derive that
\begin{eqnarray*}
	&  & \E
	\sup_{\btheta_{1} \in \bTheta_{1} }  \displaystyle\bigg\{ \frac{(y_{t} - m_{t}(\bphi))^{2}}{h_t(\omega, \balpha)} \displaystyle\bigg\}  \nonumber \\
	&= & \E
	\sup_{\btheta_{1} \in \bTheta_{1} }  \displaystyle\bigg\{ \frac{(\eps_t + m_{t}(\bphi_0) - m_{t}(\bphi))^2 }
	{h_{t}(\omega, \balpha)} \displaystyle\bigg\}
	\nonumber \\
	&\leq& \E \sup_{\btheta_{1} \in \bTheta_{1} }  \displaystyle\bigg\{ \frac{ 2 \eps_t^2 }
	{h_{t}(\omega, \balpha)} \displaystyle\bigg\} + \E \sup_{\btheta_{1} \in \bTheta_{1} }  \displaystyle\bigg\{ \frac{2( m_{t}(\bphi_0) - m_{t}(\bphi))^2 }
	{h_{t}(\omega, \balpha)} \displaystyle\bigg\} \nonumber \\
	&=& \E \sup_{\btheta_{1} \in \bTheta_{1} }  \displaystyle\bigg\{ \frac{ 2 \eta_t^2 h_t(\omega_0, \balpha_0) }
	{h_{t}(\omega, \balpha)} \displaystyle\bigg\} + \E \sup_{\btheta_{1} \in \bTheta_{1} }  \displaystyle\bigg\{ \frac{2( m_{t}(\bphi_0) - m_{t}(\bphi))^2 }
	{h_{t}(\omega, \balpha)} \displaystyle\bigg\}.
\end{eqnarray*}
The first term is bound by some constant because of $\E \eta^2 = 1$. For the second term, we have
\beq
\frac{(m_t(\bphi_0) - m_t(\bphi))^2}{h_t(\omega, \balpha)}
\leq \frac{2 \sum_{l = 1}^p (\phi_{l0} - \phi_l)^2 y_{t-l}^2}{h_t(\omega, \balpha)}
\leq \frac{C \sum_{l = 1}^p y_{t-l}^2}{\underline{\omega} + \sum_{j=1}^p \underline{\alpha_j}
	y_{t-j}^2 } < \infty.
\eeq
Then it follow that
\beq
\E \sup_{\btheta_{1} \in \bTheta_{1} }  \displaystyle\bigg\{ \frac{(y_{t} - m_{t}(\bphi))^{2}}{h_t(\omega, \balpha)} \displaystyle\bigg\} < \infty.
\eeq
Therefore,
\begin{align*}
	\E \sup_{\btheta_1 \in \bTheta_1}
	\displaystyle\bigg\{  \frac{\partial W_{t}(\btheta)} {\partial \phi_j } \displaystyle\bigg\}^{2}
	&\leq
	\E
	\sup_{\btheta \in \bTheta }
	2 C \sum ^{K}_{k=1} \frac{y_{t-j}^{2}(\widetilde y_t(\btheta_1)^{2} + \mu_{k}^{2})}{h_t(\omega, \balpha)} \\
	&\leq
	2 C \sum ^{K}_{k=1}
	\E  \sup_{\btheta \in \bTheta } \left( \frac{y_{t-j}^{2}(y_{t}- m_t(\bphi))^{2}}{h_t^2(\omega, \balpha)}
	+ \mu_{k}^{2} \frac{y^{2}_{t-j}}{h_t(\omega, \balpha)}
	\right)    \\
	&\leq
	C_{1} +
	C_{2} \sum ^{K}_{k=1}
	\E \sup_{\btheta \in \bTheta }
	\frac{y_{t-j}^{2} ( \eps_{t} + m_t(\bphi_0) - m_t(\bphi) )^{2}}
	{h_t^2(\omega, \balpha)}
	\\
	&\leq
	C_{1} +
	2 C_{2} \sum ^{K}_{k=1}
	\E \sup_{\btheta \in \bTheta }
	\frac
	{y_{t-j}^{2} [ \eps_{t}^{2} + (m_t(\bphi) - m_t(\bphi_{0}))^{2}] }
	{h_t^2(\omega, \balpha)}
	\\
	&\leq
	C_{1} +
	2 C_{2} \sum ^{K}_{k=1}
	\displaystyle\bigg ( \E
	\frac{4 y_{t-j}^2\sum_{l=1}^p \widetilde{\phi_l}^{2} y_{t-l}^{2}}{h_t^2(\underline{\omega}, \underline{\balpha})}
	+ \E
	\frac{y_{t-j}^{2} \eps^{2}_{t}}{h_t^2(\underline{\omega}, \underline{\balpha})}
	\displaystyle\bigg)   \\
	&\leq
	C_{3} + 2 C_{2} \sum ^{K}_{k=1}
	\E     \frac{y_{t-1}^{2}(\omega_{0} + \alpha_{0} y^{2}_{t-1})}{h_t^2(\underline{\omega}, \underline{\balpha})}
	\\
	&< \infty.
\end{align*}
Here $C, C_{1}, C_{2}$ and $C_3$ are several positive constants. By the similar calculation,
we can show that
\begin{align*}
	\E \sup_{\btheta \in \bTheta}
	\displaystyle\bigg\{  \frac{\partial W_{t}(\btheta)} {\partial \omega } \displaystyle\bigg\}^{2}, \quad
	\E \sup_{\btheta \in \bTheta}
	\displaystyle\bigg\{ \frac{\partial W_{t}(\btheta)} {\partial \alpha_j } \displaystyle\bigg\}^{2},
	\quad j=1, \cdots, p,
\end{align*}
\begin{align*}
	\E \sup_{\btheta \in \bTheta}
	\displaystyle\bigg\{  \frac{\partial W_{t}(\btheta)} {\partial p_{k} } \displaystyle\bigg\}^{2}, \quad
	\E \sup_{\btheta \in \bTheta}
	\displaystyle\bigg\{  \frac{\partial W_{t}(\btheta)} {\partial \mu_{k} } \displaystyle\bigg\}^{2}, \quad
	\E \sup_{\btheta \in \bTheta}
	\displaystyle\bigg\{  \frac{\partial W_{t}(\btheta)} {\partial \sigma_{k} } \displaystyle\bigg\}^{2},
\end{align*}
and the cross-products of partial derivatives of $ W_{t}(\btheta)$ with respect to parameters are all finite. This completes the proof of the first part of Lemma \ref{lem43}, equation \eqref{lem43-1}. By using the
ergodic theorem and the similar strategy in Theorem 4.2.1 in \cite{A1985}, the second part in Theorem~\ref{lem43} is derived.

\begin{lem}
	\label{lem44}
	Given Assumptions ~(A1-A6),  it follows that
	\begin{eqnarray}
		\E  \displaystyle\bigg\{ \frac{\partial W_{t}(\btheta_{0})}{\partial \btheta}
		\frac{\partial W_{t}(\btheta_{0})}{\partial \btheta^{T} } \displaystyle\bigg\}  =  \widetilde{\Omega}
	\end{eqnarray}
	is positive definite and
	\begin{eqnarray}
		\frac{1}{\sqrt{n}} \sum ^{n}_{t=p+1}  \frac{\partial W_{t}(\btheta_{0})}{\partial \btheta }  \xrightarrow{d} N (0,\widetilde{\Omega}),
	\end{eqnarray}
	where $\xrightarrow{d}$ stands by the convergence in distribution.
\end{lem}

\bigskip

\noindent\textbf{\underline{Proof:}} \ The first part of Lemma \ref{lem44} is a straightforward result of Lemma \ref{lem43}. Then by martingale central limit theorem in \cite{S1974} and Cramer-Wold theorem, we
can prove the second part of Lemma \ref{lem44}.

\begin{lem}
	\label{lem45}
	Given Assumptions ~(A1-A5), it follows that
	\begin{eqnarray}
		\E \sup_{\btheta \in \bTheta }  	\left\|\frac{\partial^{2} W_{t}(\btheta)}{\partial \btheta   {\partial \btheta^{T} } }\right\|    <\infty,
	\end{eqnarray}
	and
	\begin{eqnarray}
		\sup_{\btheta \in \bTheta }
		\left\|
			\frac{1}{n} \sum ^{n}_{t=p+1}  \displaystyle\bigg(  \frac{\partial^{2} W_{t}(\btheta)}{\partial \btheta {\partial \btheta^{T}  }} - \E  \frac{\partial^{2} W_{t}(\btheta)}{\partial \btheta {\partial \btheta^{T}   } }  \displaystyle\bigg)
		\right\|
		= o_{p}(1), \quad n \rightarrow \infty.
	\end{eqnarray}
\end{lem}

\vspace{0.1in}\

\noindent\textbf{\underline{Proof:}} \  Similar to the proof of Lemma \ref{lem43}, since that
\begin{align}
	\frac{\partial^{2} W_{t}(\btheta)}{\partial \phi_j^{2} }
	&=
	-(\sum ^{K}_{k=1} p_{k} f_{k} )^{-2}
	\displaystyle\bigg(
	\sum ^{K}_{k=1} p_{k} f_{k} \frac{y_{t-j}(\widetilde y_t(\btheta_1) - \mu_{k})}{\sigma_{k}^{2}\sqrt{h_t(\omega, \balpha)}}
	\displaystyle\bigg)  ^{2}  \nonumber   \\
	& \quad +
	(\sum ^{K}_{k=1} p_{k} f_{k} )^{-1}
	\sum ^{K}_{k=1}
	p_{k} f_{k}
	\frac{y_{t-j}^{2}(\widetilde y_t(\btheta_1)-\mu_{k})^{2} - y_{t-j}^{2}\sigma^{2}_{k}}{\sigma^{4}_{k}
		h_t(\omega, \balpha)},
\end{align}
we obtain the Lemma \ref{lem45}. Other components of the Hessian matrix are dealt by the similar way. The
second order derivatives are present as follows.
\begin{align*}
	\frac{\partial^{2} W_{t}(\btheta)}{\partial \phi_j \partial \phi_i}
	&=
	\frac{1}{\sum ^{K}_{k=1} p_{k} f_{k}}
	\sum ^{K}_{k=1}
	p_{k} f_{k}
	\frac{- y_{t-j} y_{t-i} }{\sigma^{2}_{k} h_t(\omega, \balpha)}, \quad i \neq j, \\
	\frac{\partial^{2} W_{t}(\btheta)}{\partial \phi_j \partial \alpha_i}
	&=
	\frac{1}{\sum ^{K}_{k=1} p_{k} f_{k}}
	\sum ^{K}_{k=1}
	p_{k} f_{k}
	\frac{y_{t-j} y_{t-i}^2 (-2 \widetilde y_t(\btheta_1) + \mu_k)}
	{2 \sigma^{2}_{k} h_t^{3/2}(\omega, \balpha)}, \\
	\frac{\partial^{2} W_{t}(\btheta)}{\partial \phi_j \partial \omega}
	&=
	\frac{1}{\sum ^{K}_{k=1} p_{k} f_{k}}
	\sum ^{K}_{k=1}
	p_{k} f_{k}
	\frac{y_{t-j} (-2 \widetilde y_t(\btheta_1) + \mu_k)}
	{2 \sigma^{2}_{k} h_t^{3/2}(\omega, \balpha)}, \\
	\frac{\partial^{2} W_{t}(\btheta)}{\partial \omega^{2} }
	&=\frac1{\sum ^{K}_{k=1} p_{k} f_{k} }
	\sum ^{K}_{k=1} p_{k} f_{k} \frac{\widetilde y_t(\btheta_1)( -4 \widetilde y_t(\btheta_1) + 3\mu_{k})}{ 4\sigma_{k}^{2} h_t^2(\omega, \balpha)} + \frac{1}{2 h_t^2(\omega, \balpha)},  \\
	\frac{\partial^{2} W_{t}(\btheta)}{\partial \alpha_j^{2} }
	&= \frac{y_{t-j}^4}{\sum ^{K}_{k=1} p_{k} f_{k} }
	\sum ^{K}_{k=1} p_{k} f_{k} \frac{\widetilde y_t(\btheta_1)( -4 \widetilde y_t(\btheta_1) + 3\mu_{k})}{ 4\sigma_{k}^{2} h_t^2(\omega, \balpha)} + \frac{y_{t-j}^4}{2 h_t^2(\omega, \balpha)}, \\
	\frac{\partial^{2} W_{t}(\btheta)}{\partial \alpha_j \partial \alpha_i }
	&= \frac{y_{t-j}^2 y_{t-i}^2}{\sum ^{K}_{k=1} p_{k} f_{k} }
	\sum ^{K}_{k=1} p_{k} f_{k} \frac{\widetilde y_t(\btheta_1)( -4 \widetilde y_t(\btheta_1) + 3\mu_{k})}{ 4\sigma_{k}^{2} h_t^2(\omega, \balpha)} + \frac{y_{t-j}^2  y_{t-i}^2}{2 h_t^2(\omega, \balpha)}, \quad i \neq j, \\
	\frac{\partial^{2} W_{t}(\btheta)}{\partial \alpha_j \partial \omega }
	&= \frac{y_{t-j}^2}{\sum ^{K}_{k=1} p_{k} f_{k} }
	\sum ^{K}_{k=1} p_{k} f_{k} \frac{\widetilde y_t(\btheta_1)( -4 \widetilde y_t(\btheta_1) + 3\mu_{k})}{ 4\sigma_{k}^{2} h_t^2(\omega, \balpha)} + \frac{y_{t-j}^2}{2 h_t^2(\omega, \balpha)}.
\end{align*}

\section{Proof of Theorem \ref{theorem41}}

Since the parameter space $\bTheta$ is a compact set, $\btheta_{0}$ is an inner point of $\bTheta$,
and function $l_{n}(\btheta)$ is continuous on $\bTheta$.
Then, by Lemma \ref{lem41}, we know
\begin{eqnarray*}
	l_{n}(\btheta)
	= \frac{1}{n} \sum ^{n}_{t=p+1} W_{t}(\btheta)
	\xrightarrow{P}
	\E W_{t}(\btheta), \quad \forall \btheta \in \bTheta.
\end{eqnarray*}
Furthermore, by Lemma \ref{lem42}, $\E W_{t}(\btheta)$ achieves its unique maximum at point $\btheta=\btheta_{0}$.
Combined with these results, conditions in Theorem 4.1.1 in \cite{A1985} hold and it implies that
\begin{eqnarray*}
	\widehat{\btheta}_{n} \xrightarrow{P} \btheta_{0}, \quad \mbox{as }  n \rightarrow \infty.
\end{eqnarray*}
The proof of Theorem~\ref{theorem41} is completed.

\section{Proof of Theorem \ref{theorem42}}

By Lemma \ref{lem43}, we have that
\begin{eqnarray*}
	\frac{1}{n} \sum ^{n}_{t=p+1} \frac{\partial^{2} W_{t}(\btheta)}{\partial \btheta {\partial \btheta^{T} }}.
\end{eqnarray*}
exists and is continuous in $\bTheta$. By Lemma \ref{lem45}, if the sequence of $\{\btheta_{n}\}$ satisfies
\begin{eqnarray*}
	\frac{1}{n} \sum ^{n}_{t=p+1}
	\frac{\partial^{2} W_{t}(\btheta_{n})}{\partial \btheta  {\partial \btheta^{T} }}
	\xrightarrow{P} \widetilde{\Sigma},  \quad n \rightarrow \infty,
\end{eqnarray*}
where
\begin{eqnarray*}
	\widetilde{\Sigma}
	= \E
	\frac{\partial^{2} W_{t}(\btheta_{0})}{\partial \btheta {\partial \btheta^{T}  }},
\end{eqnarray*}
then
\begin{eqnarray*}
	\btheta_{n}  \xrightarrow{P} \btheta_{0}, \quad n \rightarrow \infty.
\end{eqnarray*}
At the end, by Lemma \ref{lem44}, we have
\begin{eqnarray}
	\frac{1}{\sqrt{n}} \sum ^{n}_{t=p+1}  \frac{\partial W_{t}(\btheta_{0})}{\partial \btheta }
	\xrightarrow{d}
	N (0,\widetilde{\Omega}).
\end{eqnarray}
Finally, we verify all the conditions in Theorem 4.1.3 of \cite{A1985} and obtain that
\begin{eqnarray}
	\sqrt{n} (\hat\btheta _{n} - \btheta_{0})
	\xrightarrow{d}
	N (0, \widetilde{\Sigma}^{-1} \widetilde{\Omega} \widetilde{\Sigma}^{-1}).
\end{eqnarray}
The proof of Theorem \ref{theorem42} is completed.

\bibliographystyle{Chicago}

\begin{thebibliography}{99}
	
	\bibitem[Bai \emph{et al.}, 2003]{BRT2003}
	Bai, X., Russell, J. R. and Tiao, G. C. (2003).
	Kurtosis of GARCH and stochastic volatility models with non-normal innovations.
	\emph{Journal of Econometrics}, $\mathbf{114}$, 349--360.
	
	\bibitem[Baudry \emph{et al.}, 2012]{BMM2012}
	Baudry, J.-P., Maugis, C. and Michel, B. (2012).
	Slope heuristics: overview and implementation.
	\emph{Stat. Comput.}, \textbf{22}, 455--470.
	
	\bibitem[Berkes and Horvath, 2004]{BH2004}
	Berkes, I. and Horvath, L. (2004).
	The efficiency of the estimators of the parameters in GARCH processes.
	\emph{Ann. Statist.}, $\mathbf{32}$, 633--655.
	
	\bibitem[Berkes \emph{et al.}, 2003]{BHK2003}
	Berkes, I., Horvath, L. and Kokoszka, P. (2003).
	GARCH processes: structure and estimation.
	\emph{Bernoulli}, $\mathbf{9}$, 201--207.
	
	\bibitem[Biernacki \emph{et al.}, 2002]{BCG2002}
	Biernacki, C., Celeux, G. and Govaert, G. (2002).
	Assessing a mixture model for clustering with the integrated completed likelihood.
	\emph{IEEE Trans. Pattern Anal. Mach. Intell.}, \textbf{22}(7), 719--725.
	
	\bibitem[Birgé and Massart, 2007]{BM2007}
	Birgé, L. and Massart, P. (2007).
	Minimal penalties for Gaussian model selection.
	\emph{Probab. Theory Relat. Fields}, \textbf{138}, 33--73.
	
	\bibitem[Bollerslev, 1987]{B1987}
	Bollerslev, T. (1987).
	A conditionally heteroskedastic time series model for speculative prices and rates of return.
	\emph{Rev. Econ. Statist}, $\mathbf{69}$, 542--547.
	
	\bibitem[Borkovec and Kluppelberg, 1998]{BK1998}
	Borkovec, M. and Kluppelberg, C. (1998).
	The tail of the stationary distribution of an autoregressive process with ARCH(1) errors.
	\emph{Ann. Appl. Probab.}, $\mathbf{11}$, 1220--1241.
	
	\bibitem[Christofersen, 1998]{C1998}
	Christofersen, P. F. (1998).
	Evaluating Interval Forecasts.
	\emph{International Economic Review}, \textbf{39}(4), 841--862.
	
	\bibitem[Fiorentini and Sentana, 2007]{fiorentini2007efficiency}
	Fiorentini, G. and Sentana, E. (2007).
	On the efficiency and consistency of likelihood estimation in multivariate conditionally heteroskedastic dynamic regression models.
	CEMFI Working Paper 0713.
	
	\bibitem[Fiorentini and Sentana, 2019]{fiorentini2019consistent}
	Fiorentini, G. and Sentana, E. (2019).
	Consistent non-Gaussian pseudo maximum likelihood estimators.
	\emph{Journal of Econometrics}, \textbf{213}(2), 321--358.
	
	\bibitem[Francq and Zako\"{l}an, 1998]{FZ1998}
	Francq, C. and Zako\"{l}an, J. M. (1998).
	Estimating linear representations of nonlinear processes.
	\emph{J. Statist. Plann. Inference}, $\mathbf{68}$, 145--165.
	
	\bibitem[Francq and Zako\"{l}an, 2000]{FZ2000}
	Francq, C. and Zako\"{l}an, J. M. (2000).
	Covariance matrix estimation for estimators of mixing weak ARMA models.
	\emph{J. Statist. Plann. Inference}, $\mathbf{83}$, 369--394.
	
	\bibitem[Francq and Zako\"{l}an, 2010]{FZ2010}
	Francq, C. and Zako\"{l}an, J. M. (2010).
	\emph{GARCH Models: Structure, Statistical Inference and Financial Applications.}
	John Wiley.
	
	\bibitem[Ghorbanzadeh \emph{et al.}, 2014]{GLD2014}
	Ghorbanzadeh, D., Luan, J., and Durand, P. (2014).
	A method to simulate the skew normal distribution.
	\emph{Applied Mathematics}, $\mathbf{5}(13)$, 2073--2076.
	
	\bibitem[Gong and Li, 2020]{GL2020}
	Gong, H., and Li, D. (2020).
	On the three-step non-Gaussian quasi-maximum likelihood estimation of heavy-tailed double autoregressive models.
	\emph{Journal of Time Series Analysis}, $\mathbf{41}$(6), 883--891.
	
	\bibitem[Guegan and Diebolt, 1994]{GD1994}
	Guegan, D. and Diebolt, J. (1994).
	Probabilistic properties of the $\beta$-ARCH-model.
	\emph{Statist. Sin.}, $\mathbf{7}$, 71--87.
	
	\bibitem[Ha and Lee, 2011]{HL2011}
	Ha, J., and Lee, T. (2011).
	NM-QELE for ARMA-GARCH models with non-Gaussian innovations.
	\emph{Statistics and Probability Letters}, $\mathbf{81}(6)$, 694--703.
	
	\bibitem[Haas \emph{et al.}, 2004]{HMP2004}
	Haas, M., Mittnik, S. and Paolella, M. S. (2004).
	Mixed normal conditional heteroskedasticity.
	\emph{Financ. Econom}, $\mathbf{2}$, 211--250.
	
	\bibitem[Hall and Yao, 2003]{HY2003}
	Hall, P. and Yao, Q. (2003).
	Inference in ARCH and GARCH models with heavy-tailed errors.
	\emph{Econometrica}, $\mathbf{71}$, 285--317.
	
	\bibitem[Keribin, 2000]{K2000}
	Keribin, C. (2000).
	Consistent estimation of the order of mixture models.
	\emph{Sankhyā Ser. A}, \textbf{62}, 49--66.
	
	\bibitem[Kupiec, 1995]{K1995}
	Kupiec, P. H. (1995).
	Techniques for Verifying the Accuracy of Risk Measurement Models.
	\emph{The Journal of Derivatives}, $\mathbf{3}(2)$, 73--84.
	
	\bibitem[Lazar and Alexander, 2004]{LA2004}
	Lazar, E. and Alexander, C. (2004).
	Symmetric normal mixture GARCH.
	\emph{EFMA 2004 Basel Meetings Paper}, SSRN: http://ssrn.com/abstract=497524.
	
	\bibitem[Lee and Lee, 2009]{LL2009}
	Lee, T. and Lee, S. (2009).
	Normal Mixture Quasi-maximum Likelihood Estimator for GARCH Models.
	\emph{Scandinavian Journal of Statistics}, $\mathbf{36}$, 157--170.
	
	\bibitem[Ling, 2004]{L2004}
	Ling, S. (2004).
	Estimation and testing stationarity for double autoregressive models.
	\emph{J. Roy. Statist. Soc. Ser. B}, $\mathbf{66}$, 63--78.
	
	\bibitem[Ling, 2007]{L2007}
	Ling, S. (2007).
	A double AR(p) model: structure and estimation.
	\emph{Statistica Sinica}, $\mathbf{17}$, 161--175.
	
	\bibitem[Ling and McAleer, 2003]{LM2003}
	Ling, S. and McAleer, M. (2003).
	Asymptotic theory for a new vector ARMA-GARCH model.
	\emph{Econometric Theory}, $\mathbf{19}$, 280--310.
	
	
	\bibitem[McLachlan and Peel, 2000]{MP2000}
	McLachlan, G. J. and Peel, D. (2000).
	\emph{Finite Mixture Models}.
	John Wiley \& Sons.
	
	\bibitem[McLachlan and Rathnayake, 2014]{MR2014}
	McLachlan, G. J. and Rathnayake, S. (2014).
	On the number of components in a Gaussian mixture model.
	\emph{Wiley Interdiscip. Rev. Data Min. Knowl. Discov.}, \textbf{4}(5), 341--355.
	
	\bibitem[McLachlan \emph{et al.}, 2019]{MLR2019}
	McLachlan, G. J., Lee, S. X. and Rathnayake, S. I. (2019).
	Finite mixture models.
	\emph{Annu. Rev. Stat. Appl.}, \textbf{6}(1), 355--378.
	
	\bibitem[Mikosch and Starica, 2000]{MS2000}
	Mikosch, T. and Starica, C. (2000).
	Limit theory for the sample autocorrelations and extremes of a GARCH(1,1) process.
	\emph{Ann. Statist.}, $\textbf{28}$, 1427--1451.
	
	\bibitem[Newey and Steigerwald, 1997]{newey1997asymptotic}
	Newey, W. K. and Steigerwald, D. G. (1997).
	Asymptotic bias for quasi-maximum-likelihood estimators in conditional heteroskedasticity models.
	\emph{Econometrica}, \textbf{65}, 587--599.
	
	\bibitem[Redner, 1984]{R1984}
	Redner, R. (1984).
	Notes on the consistency of the maximum likelihood estimate for nonidentifiable distributions.
	\emph{The Annals of Statistics}, $\textbf{9}$, 225--239.
	
	\bibitem[Roberts, 2001]{R2001}
	Roberts, M. C. (2001).
	Commodities, options, and volatility: modelling agricultural futures prices.
	\emph{Working paper, Ohio State University}.
	
	\bibitem[Schwarz, 1978]{S1978}
	Schwarz, G. (1978).
	Estimating the dimension of a model.
	\emph{Ann. Statist.}, \textbf{6}, 461--464.
	
	\bibitem[Wang and Pan, 2014]{WP2014}
	Wang, H. and Pan, J. Z. (2014).
	Restricted normal mixture QMLE for non-stationary TGARCH(1, 1) models.
	\emph{Science China Mathematics}, $\mathbf{57}$, 1341--1360.
	
	\bibitem[Weiss, 1986]{W1986}
	Weiss, A. A. (1986).
	Asymptotic theory for ARCH models: estimation and testing.
	\emph{Econometric Theory}, $\mathbf{2}$, 107--131.
	
	\bibitem[Zhang \emph{et al.}, 2006]{ZLY2006}
	Zhang, Z., Li, W. K. and Yuen, K. C. (2006).
	On a mixture GARCH time-series model.
	\emph{Time Ser. Anal}, $\mathbf{27}$, 577--597.
	
	\bibitem[Zhu and Ling, 2011]{ZL2011}
	Zhu, K. and Ling, S. (2011).
	Global self-weighted and local quasi-maximum exponential likelihood estimators for ARMA-GARCH/IGARCH models.
	\emph{The Annals of Statistics}, \textbf{39}, 2131--2163.
	
	\bibitem[Zhu and Ling, 2013]{ZL2013}
	Zhu, K. and Ling, S. (2013).
	Quasi-maximum exponential likelihood estimators for a double AR (p) model.
	\emph{Statist. Sinica.}, $\mathbf{23}$, 251--270.
	
\end{thebibliography}

\begin{thebibliography}{99}
	
	\bibitem[Amemiya, 1985]{A1985}
	Amemiya, T. (1985). \emph{Advanced Econometrics}. Harvard University Press, Cambridge.
		
	\bibitem[Ling, 2004]{L2004} Ling, S. (2004).
	Estimation and testing stationarity for double autoregressive models.
	\emph{J. Roy. Statist. Soc. Ser. B},
	$\mathbf{66}$, 63--78.
	
	\bibitem[Stout, 1974]{S1974}
	Stout, W.F.(1974).
	\emph{Almost Sure Convergence.}
	New York: Academic Press.

	
\end{thebibliography}

\end{document}